\documentclass[%
 aip,
 pof,%
 amsmath,amssymb,
preprint,%
]{revtex4-1}

\usepackage{graphicx}
\usepackage{dcolumn}
\usepackage{bm}

\begin{document}

\preprint{AIP/123-QED}

\title{Curvature Singularity in the Asymmetric Breakup of an Underwater Air Bubble}

\author{Lipeng Lai}
\email{lplai@uchicago.edu.} 
\affiliation{Department of Physics and James Franck Institute, University of Chicago, Chicago, 60637, USA}
 
\date{\today}

\begin{abstract}
The presence of slight azimuthal asymmetry in the initial shape of an underwater bubble entirely alters the final breakup dynamics. Here I examine the influence of initial asymmetry on the final breakup by simulating the bubble surface evolution as a Hamiltonian evolution corresponding to an inviscid, two-dimensional, planar implosion. I find two types of breakups: a previously reported coalescence mode in which distant regions along the air-water surface curve inwards and eventually collide with finite speed, and a hitherto unknown cusp-like mode in which the surface develops sharp tips whose radii of curvature are much smaller than the average neck radius.  I present three sets of results that characterize the nature of this cusp mode.  First, I show that the cusp mode corresponds to a saddle-node. In other words, an evolution towards a cross-section shape with sharp tips invariably
later evolves away from it.  In phase space, this saddle-node separates coalescence modes whose coalescence planes lie along
different spatial orientations. Second, I show that the formation of the sharp tips can be interpreted as a weakly first-order
transition which becomes second-order, corresponding to the formation of a finite-time curvature singularity, in the limit that the initial
perturbation amplitude approaches zero. Third I show that, as the curvature singularity is approached, the maximum surface curvature diverges approximately as $(t_c - t)^{-0.8}$, where $t_c$ is the onset time of the singularity and the maximum velocity
diverges approximately as $(t_c-t)^{-0.4}$. In practice, these divergences imply that viscous drag and compressibility of the gas flow, two effects not included in my analysis, become significant as the interface evolves towards the curvature singularity.
\end{abstract}

\maketitle

\section{\label{section:introduction}Introduction}
An underwater air  bubble breakup occurs in a wide range of natural and industrial processes: whenever a stone is thrown into a pond, rain drops hit the ocean's surfaces, or a propeller is running with its maximum power, gas cavities are produced. Those cavities often break up into smaller pieces subsequently. If the bubble neck is axisymmetric, as the neck thins down, it disconnects at one single point and forms a finite time focusing singularity  \cite{higgins91, oguz93, gordillo05, bergmann06, thoroddsen07, eggers07, gekel09}, during which relevant physical quantities including velocity and pressure diverge. Based on the argument that the length and time scales near the singularity are dramatically different from those set by initial and boundary conditions, people used to think such a breakup singularity to be universal, one independent of initial and boundary conditions. This idea successfully describes several breakup scenarios, such as water drops break up in air \cite{eggers94, shi94, chen97, eggers97, chen02}. However, recent studies reported that there are breakups which remember their initial states \cite{doshi03, burton05, keim06, schmidt09, turitsyn09, keim11, enriquez10, enriquez11, enriquez11_arxiv}.

\begin{figure}
\begin{center}
\includegraphics[width=13.5cm]{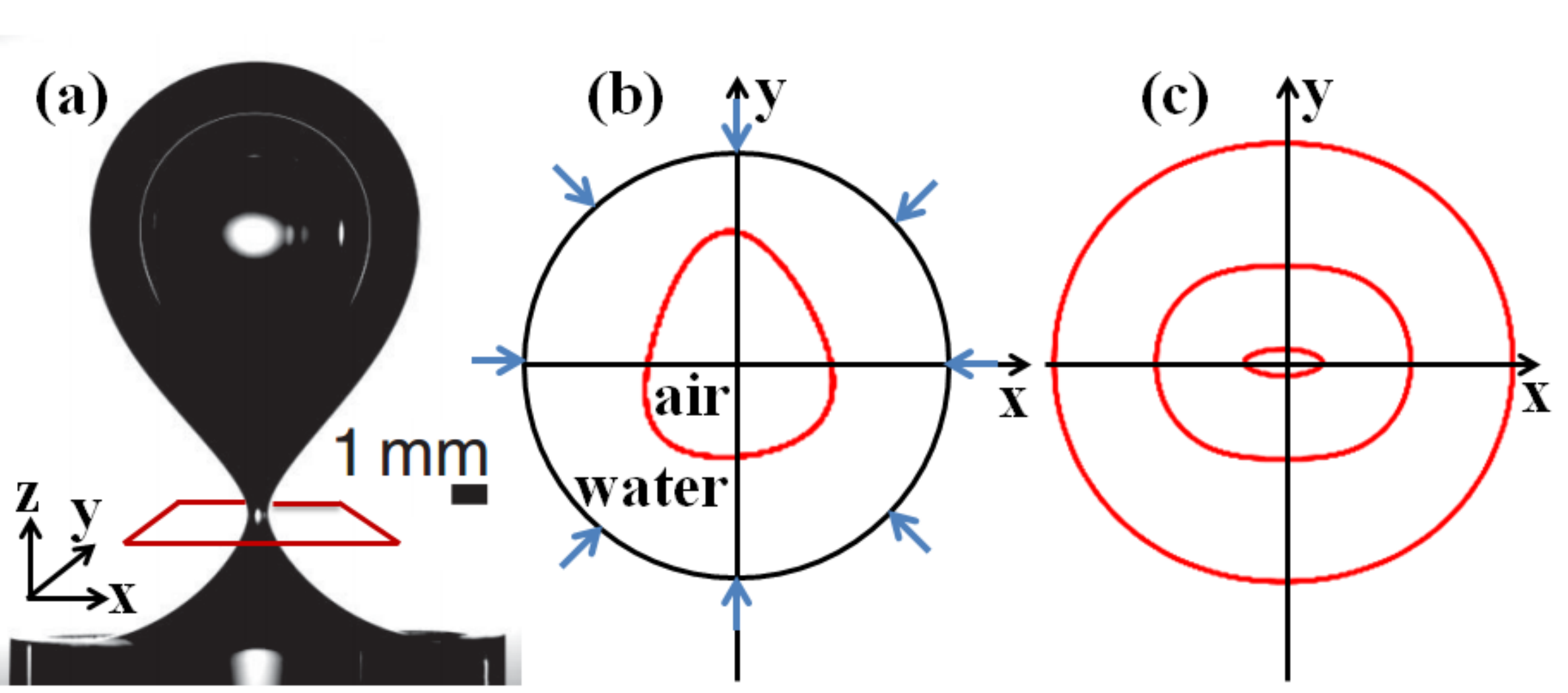}
\caption{\label{fig:setup}Bubble breakup dynamics. (a) Experiment: an underwater air bubble (dark area) is released from a nozzle. Bright spots are optical artifacts. (Image courtesy of N. C. Keim and S. R. Nagel.) (b) Model: two dimensional horizontal cross section of the bubble neck. The air-water interface is represented by the red distorted curve that contracts due to the radial flow at far field. (c) Cross-section shape sequence of the contraction in laboratory frame. Each closed curve represents the air-water interface at a certain time. It shows that starting with slight perturbations (outer-most curve), the interface evolves into an elongated shape with sharp tips at its east and west ends (inner-most curve).}
\end{center}
\end{figure}

\begin{figure*}
\begin{center}
\includegraphics[width=16cm]{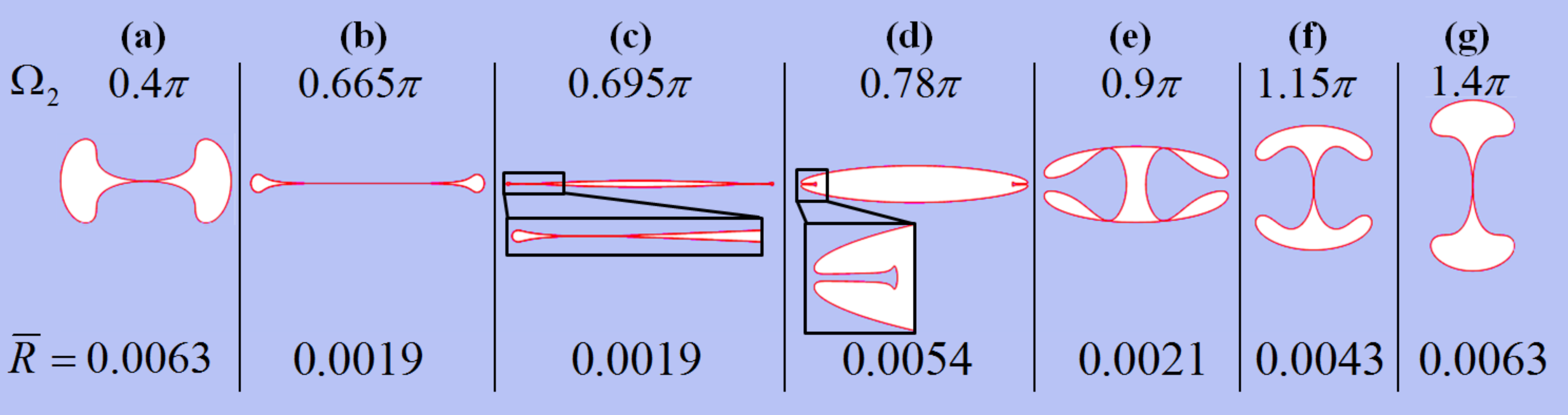}
\caption{\label{fig:shapes}Different initial phases (with initial amplitude $A_2=0.01$ fixed) give rise to different modes of breakups in simulations. Bubble cross-section shapes near breakups are shown here for different initial phases $\Omega_2$. The air-water interfaces are indicated by red curves and the average radius $\bar{R}$ for each cross-section is labeled below each corresponding cross-section shape. For two initial phases different by $\pi$, outcomes are identical up to a rotation by 90$^\circ$ as can be seen from (a) $\Omega_2=0.4\pi$ and (g) $\Omega_2=1.4\pi$. For a wide range of initial phases, the interface evolves into a one-point coalescence (e.g., (a) $\Omega_2=0.4\pi$). For a different range of initial phases, the interface develops sharp tips from which re-entrant water fingers form (e.g., (d) $\Omega_2=0.78\pi$). In the transition from one breakup mode (one-point coalescence) to the other (interface with sharp tips), the interface ends up with a multiple-point coalescence (e.g., (c) $\Omega_2=0.695\pi$ and (e) $\Omega_2=0.9\pi$). As the initial phase decreases towards around 0.7$\pi$, the east and west ends of the cross-section shape reach higher and higher curvatures before the formation of re-entrant water fingers.}
\end{center}
\end{figure*}

One example I am looking at here is the breakup of an underwater air bubble (Fig. \ref{fig:setup} (a)). Keim \emph{et al.} \cite{keim06} found in experiments that any initial slight asymmetry destroys the symmetric focusing singularity and produces different forms of breakup depending on initial conditions. Later, Schmidt \emph{et al.} \cite{schmidt09} showed analytically that the focusing singularity is pre-empted by persistent standing waves in forms of azimuthal vibrations along the bubble neck excited by initial azimuthal asymmetries. Those vibrations encode a memory of initial states, and are thought to be possibly one common feature of focusing singularities \cite{schmidt09, whitham57, plesset77, schmidt11}. The combination of singular dynamics and wave dynamics reveals the interesting but challenging aspect of such a problem: On the one hand, in linear stability analysis \cite{schmidt09}, those waves excited by initial azimuthal asymmetries preserve their amplitudes while their vibrational frequencies chirp (vibrational frequencies diverge as the area of the neck cross-section shrinks to zero). On the other, the bubble neck evolves towards a breakup singularity with ever shrinking cross-section. Then it is immediately noticed that, no matter how small the vibrational amplitude is initially, it will become significant when the average size of the neck cross-section decreases to a size comparable to the amplitude (e.g., Fig. \ref{fig:setup} (c)).  As a consequence, the dynamics of the bubble breakup inevitably evolves into a nonlinear region. Turitsyn \emph{et al.} \cite{turitsyn09} studied consequences of the memory when the dynamics becomes nonlinear which cannot be addresses in the linear theory by Schmidt \emph{et al.} \cite{schmidt09}. They showed that one generic outcome due to the persistent neck vibrations is that instead of the symmetric focusing singularity, the neck cross-section evolves into a smooth contact/coalescence, in which distant points along the air-water interface touch each other with finite speed. Their results are also reproduced by my simulations and one example of the coalescence can be seen in Fig. \ref{fig:shapes} (a) ($\Omega_2=0.4\pi$). In this case, the nonlinearity is shown to be weak in the sense that the time evolution of the interface is well approximated by linear dynamics after a set of appropriate nonlinear transformations \cite{turitsyn09, shraiman84, dyachenko96, zakharov02}. However, interfaces that develop high curvature regions and then become re-entrant were not studied by their simulations using spectral method. This is mainly because the previous method is based on conformal mapping, which limits the spatial resolution in simulations when the interface has re-entrant regions.

There are evidences \cite{keim06, turitsyn09, enriquez10, enriquez11, enriquez11_arxiv} supporting that for some initial conditions, the interface develops high curvature regions followed by re-entrant water fingers. Using simulations with boundary integral method, I found that, starting with one single mode perturbation, there is a continuous range of initial conditions, for which the air-water interface develops convex high curvature regions -- sharp tips. For each of these high curvature regions, it evolves into a sharper and sharper tip. After the curvature at the tip reaches an extreme value, the tip flattens, reverses its sign of curvature and develops a re-entrant water finger intruding into the air. This first-sharpen-then-flatten tip evolution is qualitatively consistent with a phase space trajectory controlled by a saddle-node. The extreme tip curvature appears to diverge when initial amplitude and phase are tuned towards appropriate values. This suggests that the saddle-node corresponds to a curvature singularity. Formations of curvature singularities with free surface flows are of general mathematical and physical interest \cite{shraiman84, bensimon86, tanveer93, tanveer95, siegel00, lee06}. The question I am interested in here is whether there is an initial condition (or a family of initial conditions) for which the symmetric breakup singularity is replaced by this curvature singularity. Usually, such a curvature singularity is regularized by different physical effects, such as surface tension and viscosity. Within the model described in Section \ref{section:problem}, I will present my simulation results showing that this curvature singularity in the asymmetric bubble breakup can be cut off just due to the collapsing dynamics. More specifically, I found that the interface appears to evolve into the curvature singularity by developing the sharp tip with infinite curvature when the initial condition is tuned towards some threshold value. However, starting with the initial condition at the threshold value, the interface evolves into a coalescence before the tip curvature diverges. Thus this curvature singularity is pre-empted by coalescence and cannot be realized. The numerics further suggest that the curvature singularity can only be reached with vanishingly small perturbation amplitude.

This paper is structured as follows. In Section \ref{section:problem}, I will setup the physical system I am looking at and describe associated equations. After that, in Section \ref{section:method}, I will elaborate on my numerical approach to this problem. Then in Section \ref{section:result}, I will start with describing the interface evolution towards and after the formation of sharp tips. Following that, I will focus on two aspects of the curvature singularity. I will first discuss the phase space behavior around the singularity, and then for a given phase space trajectory, I will show the scalings of relevant quantities when the singularity is approached. Finally, in Section \ref{section:discussion}, I will first discuss the physical effects from surface tension, viscosity and air compressibility, which are currently ignored in my model. I will show that both air viscosity and compressibility will regularize the curvature singularity if it were to form in experiments. Then I will briefly discuss the connections between my numerical results and experiments.

\section{\label{section:problem}Problem Formulation}
Previous studies of air bubble breakup showed that when the average radius of the horizontal cross-section of the bubble neck minimum approaches zero, the bubble neck can be well approximated by a long and slender shape \cite{bergmann06, schmidt11, herbst11}. In this asymptotic limit, the dynamics of water at different heights are decoupled from each other. The horizontal cross-sections around the neck minimum shrink and deform only due to the dynamics of water in the same 2D plane. I use the same approach as the one used in previous studies \cite{schmidt09, turitsyn09}, which focuses on the two dimensional dynamics within the horizontal plane crossing the neck minimum, where the breakup first happens (Fig. \ref{fig:setup} (b)).

For the average (axisymmetric) collapse of the bubble neck, the water inertia dominates viscosity, surface tension and gravity as the neck radius approaches zero \cite{schmidt09}. In my model, I ignore those terms. The exterior fluid, water, is described by a velocity field ${\bf u}$ that is both incompressible ($\nabla\cdot\bf{u}=0$) and irrotational ($\bf{\nabla}\times\bf{u}=0$). This then allows the velocity field to be described by a velocity potential $\phi({\bf{x}},t)$, a scalar field satisfying the Laplace's equation $\Delta\phi=0$, and such that $\bf{u}=\bf{\nabla}\phi$. I also ignore the dynamics of air in interior. The air pressure $p_{air}(t)$ in the bubble is assumed to be uniform, whose value ensures prescribed implosion areal flux from infinity.

On the interface between water and air, the dynamical boundary condition states the balance between relevant stresses. This gives a differential equation for the time evolution of the velocity potential $\phi$:
\begin{equation}
\rho(\frac{D\phi}{D t}-\frac{1}{2}|{\bf{\nabla}}\phi|^2)|_S=-p_{air}
\label{bc:dynamic}
\end{equation}
where $\rho$ is the density of water, and $\frac{D\phi}{Dt}=\frac{\partial\phi}{\partial t}+{\bf{u}\cdot\bf{\nabla}}\phi$ is the material derivative taken along a path co-moving with $\bf{u}$.
The kinematic boundary condition states that the fluid particles on the interface are advected by the velocity at their field points:
\begin{equation}
\frac{D{\bf{x}}}{Dt}|_S=(\frac{\partial}{\partial t}+{\bf{u}}\cdot{\bf{\nabla}}){\bf{x}}={\bf{\nabla}}\phi|_S
\label{bc:kinematic}
\end{equation}
Both equation (\ref{bc:dynamic}) and (\ref{bc:kinematic}) are evaluated on the air-water interface $S$. The collapse of the neck cross-section is driven by a prescribed radial flux at far field ${\bf{u_r}}=(-Q(t)/r)\bf{e_r}$ (Fig. \ref{fig:setup} (b)), where $Q(t)$ is assumed a constant $Q$ in time in my simulations. The average radius $\bar{R}$(t) of the cross-section (defined as the radius of a circle having the same area as the cross-section) decreases as $R_0\sqrt{(t_*-t)/t_*}$, where $t_*=R_0^2/2Q$ is the time for the area of the cross-section to shrink to zero.

Estimated from the experiment of the symmetric breakup of an underwater air bubble with a $4$-$mm$-diameter circular nozzle \cite{keim11, schmidt11}, the typical initial length scale for the cross-section size is $R_0=250\mu m$ and the initial velocity scale is $u_0=0.5m/s$.

I focus on the simplest initial conditions that include only one single mode perturbation. At $t=0$, when the perturbation amplitude $A_n$ non-dimensionalized by the average size $R_0$ of the cross-section is much smaller than $1$, the initial cross-section shape is nearly a circle, described as 
\begin{equation}
S(\theta, t=0)=R_0[1+A_n\cos(\Omega_n)\cos(n\theta)]
\label{eqn:bcS}
\end{equation}
in polar coordinates, whose origin coincides with the point sink of the influx at infinity. According to the calculation by Schmidt \emph{et al.} \cite{schmidt09}, to the first order in $A_n$, the initial velocity potential $\phi$ along the interface is then described as
\begin{equation}
\phi(\theta, t=0)=-Q[\ln(R_0)-\frac{1}{n}(\sqrt{n-1}A_n\sin(\Omega_n)+(1-n)A_n\cos(\Omega_n))\cos(n\theta)]
\label{eqn:bcphi}
\end{equation}
In the rest part of this paper, unless otherwise pointed out specifically, the initial perturbation is restricted to the simplest case with only $n=2$ mode perturbation as was done in the analysis by Turitsyn \emph{et al.} \cite{turitsyn09}. However, instead of holding the initial phase $\Omega_2$ fixed and changing the amplitude $A_2$, my approach is that I first fix the initial amplitude $A_2$ and then focus on the case when initial phase $\Omega_2$ varies. This approach allows me to observe how the dynamics varies with initial condition without large changes in the dynamic range of the average radius $\bar{R}$. Finally, given the initial amplitude, for two initial phases different by $\pi$, the interface evolutions are identical up to a rotation by $90^\circ$. Thus I only focus on initial phases with a span of $\pi$.

From here, unless otherwise pointed out, all quantities are understood as being non-dimensionalized by initial length scale $R_0$, initial velocity scale $u_0$, water density $\rho$, and their combinations.

\section{\label{section:method}Numerical simulations with boundary integral method}

A numerical simulation with boundary integral method is implemented to solve the time evolution of the cross-section shape. The boundary integral method provides an efficient way to solve incompressible and irrotational flows \cite{higgins76, pozrikidis_book}, which in my case allows me to solve for $\bf{u}$ in the exterior fluid given the potential on the boundary. This method is based on the following Green's function formulation that associates the velocity potential with the normal velocity on the air-water interface:
\begin{equation}\label{integraleqn}
\frac{1}{2}\phi({\bf{x}_0})=-\oint_{{\bf{x}}\in S}G({\bf{x}_0}, {\bf{x}}){\bf{n}}\cdot{\bf{\nabla}}\phi({\bf{x}})ds({\bf{x}})+ \oint_{{\bf{x}}\in S}\phi({\bf{x}}){\bf{n}}\cdot{\bf{\nabla}}G({\bf{x}_0}, {\bf{x}})ds({\bf{x}})
\end{equation}
where the two integrals on the right hand side are along the air-water interface $S({\bf{x}}, t)$, ${\bf{x}_0}$ and ${\bf{x}}$ are both evaluated on $S$, $ds({\bf{x}})$ is the line element at ${\bf{x}}$ and ${\bf{n}}$ is the unit surface normal at ${\bf{x}}$ pointing into water. $G({\bf{x}_0}, {\bf{x}})=-\ln(|{\bf{x}_0}-{\bf{x}}|)/2\pi$ is the free-space Green's function for Laplace's equation in 2D, and ${\bf{n}}\cdot{\bf{\nabla}}\phi({\bf{x}}) = u_\perp({\bf{x}})$ gives the normal velocity $u_\perp$ at ${\bf{x}}$. Equation (\ref{integraleqn}) together with the stress balance condition (\ref{bc:dynamic}) and the kinematic boundary condition (\ref{bc:kinematic}) specifies the evolution of the interface.

To implement the numerics, I first discretize the air-water interface (starting as a simple closed curve) into $N$ boundary elements $E_i$ ($i = 1, ..., N$) separated by $N$ nodal points $\bf{x_i}$ ($i=1, ... , N$) with an adaptive mesh (necessary to provide a good nodal points distribution in resolving large curvatures and re-entrant water fingers). How to choose an appropriate adaptive meshing scheme is a tricky problem here. When the interface develops sharp tips, the tangential velocity around those convex high curvature regions tends to advect any perturbations of the interface towards the tips. This flow will potentially amplify any numerical perturbation introduced by a non-proper redistribution scheme and make further investigations of the sharp tips impossible. However, introducing a smoothing method at each time step (or after every certain number of them) also erases or alters the structures with small length scales one may be interested in. After experimenting with different redistribution schemes including moving nodal points according to curvature, I found that the stable redistribution scheme, which I used in my simulations, is actually rather simple: advecting nodal point $\bf{x_i}$ with the total velocity at the point $\bf x_i$ naturally concentrates points into regions around the sharp tips that are both convex and highly curved, providing good spatial resolution. In addition, part of the calculation procedure mentioned in the Appendix (switching between nodal points $\to$ middle points of line elements $\to$ nodal points \cite{pozrikidis_book}) also helps stabilize the interface.

To describe the initial condition of my simulation, I use the same variables as employed by Turitsyn \emph{et al.} \cite{turitsyn09}. First, the unit circle $w=e^{i\beta}$ on the complex plane is transformed onto the air-water interface in real space in 2D polar coordinates $S(r, \theta, t)$, with the exterior of the unit circle mapped conformally to the exterior of the air-water interface. Thus each point $(x, y)$ in real space in water is represented as  $z=x+iy=z(w, t)$. Then a second transformation used by Turitsyn \emph{et al.} \cite{turitsyn09} is performed defining two new variables:
\begin{equation}\label{RVdefine}
{\cal R}=\frac{1}{w\partial_wz},\quad\quad{\cal V} =\frac{\partial_w\Psi}{\partial_wz}
\end{equation}
where $\Psi(w, t)=\phi(w,t)+i\psi(w,t)$ is the complex velocity potential for the exterior fluid and $\psi(w,t)$ is the stream function. $\partial_w$ is the partial derivative with respect to $w$. The variable $\cal R$ is related to the Jacobian of the conformal mapping and it encodes the distortion to the cross-section shape, and the variable $\cal V$ represents the velocity field explicitly as ${\cal V}=u_x-iu_y$ with $u_x$ and $u_y$ the x-component and y-component of the velocity respectively. In terms of ${\cal R}$ and ${\cal V}$, the initial condition is written as the following expansions
\begin{eqnarray}
{\cal R}&=&\frac{1}{w}(1+\sum_{n>1}(n-1)A_n\cos (\Omega_n)\frac{1}{w^n})\label{eqn:Rexpansion}\\
{\cal V}&=&-\frac{1}{w}(1+\sum_{n>1}\sqrt{n-1}A_n\sin (\Omega_n)\frac{1}{w^n})\label{eqn:Vexpansion}
\end{eqnarray}
Just to restate it here, all the relevant quantities are non-dimensionalized using initial length scale $R_0$ and velocity scale $u_0$. With the expansions, the initial condition is specified by the amplitude $A_n$ and phase $\Omega_n$ of each wave mode $n$ ($n>1$). When $A_n\ll 1$, the expansions in $\cal R$ and $\cal V$ agree with both the initial condition introduced in Section \ref{section:problem} and that used in the analysis by Schmidt \emph{et al.} \cite{schmidt09, schmidt11}, to the first order in $A_n$.\\
\indent To advance in time, an explicit 4th order Runge-Kutta time-stepping scheme with variable time step sizes is used. According to the analysis by Schmidt \emph{et al.} \cite{schmidt09}, the period $T$ of the azimuthal vibrations scales as the square of the length scale of the cross-section. To  resolve such vibrations in the wave dynamics, the time-step size $\Delta t$ is prescribed to be $C |\Delta x_{min}|^2$ where $C$ (typically around $200$ to $400$ for $800$ grid points) is chosen empirically and $\Delta x_{min}$ is the minimal distance between adjacent nodal points (minimal length of elements).\\
\indent The simulation tracks the time evolution of the interface until: (1) The first topological change of the interface, numerically defined as $min\{d_{i,j}\} < d_{limit}$ where $d_{i,j}$ is the distance between the nodal point $\bf{x_i}$ and the element $E_j$, and $d_{limit}$ is a pre-calculated threshold value below which a self-contact of the interface is determined to have happened; or (2) the ratio of the local radius of curvature along the interface to the local grid spacing drops below a prescribed value, 0.5 here. The local grid spacing at point $\bf{x}_i$ is defined as the arithmetic average of its adjacent elements' lengths. More details of the numerical implementation can be found in the Appendix.\\
\indent A standard simulation uses $800$ grid points and $C=400$ for time stepping. For some initial conditions, when the interface is close to a topological change or when the curvature changes rapidly, $C$ is reduced to $100$ or $50$ during the simulations to obtain more data in respective time windows. When the initial condition only includes one single mode $n$ perturbation, an $n$-fold spatial symmetry is imposed to the system to simplify calculations. For example, when the initial perturbation only includes an $n=2$ mode, which is the case I will mainly focus on in the following sections, the cross-section remains symmetric over time with respect to the horizontal or vertical axis passing through its center. For all initial conditions I have checked, doubling the grid points ($N=1600$) or halving the time step size ($C=200$) doesn't change the results presented in this paper. The results here are also compared with linear stability analysis \cite{schmidt09} and simulations using spectral method \cite{turitsyn09} where the comparison is applicable. They are all consistent with each other quantitatively.

\section{\label{section:result}Results}
\subsection{\label{subsection:phenomena}Phenomena}
\begin{figure}
\begin{center}
\includegraphics[width=13.5cm]{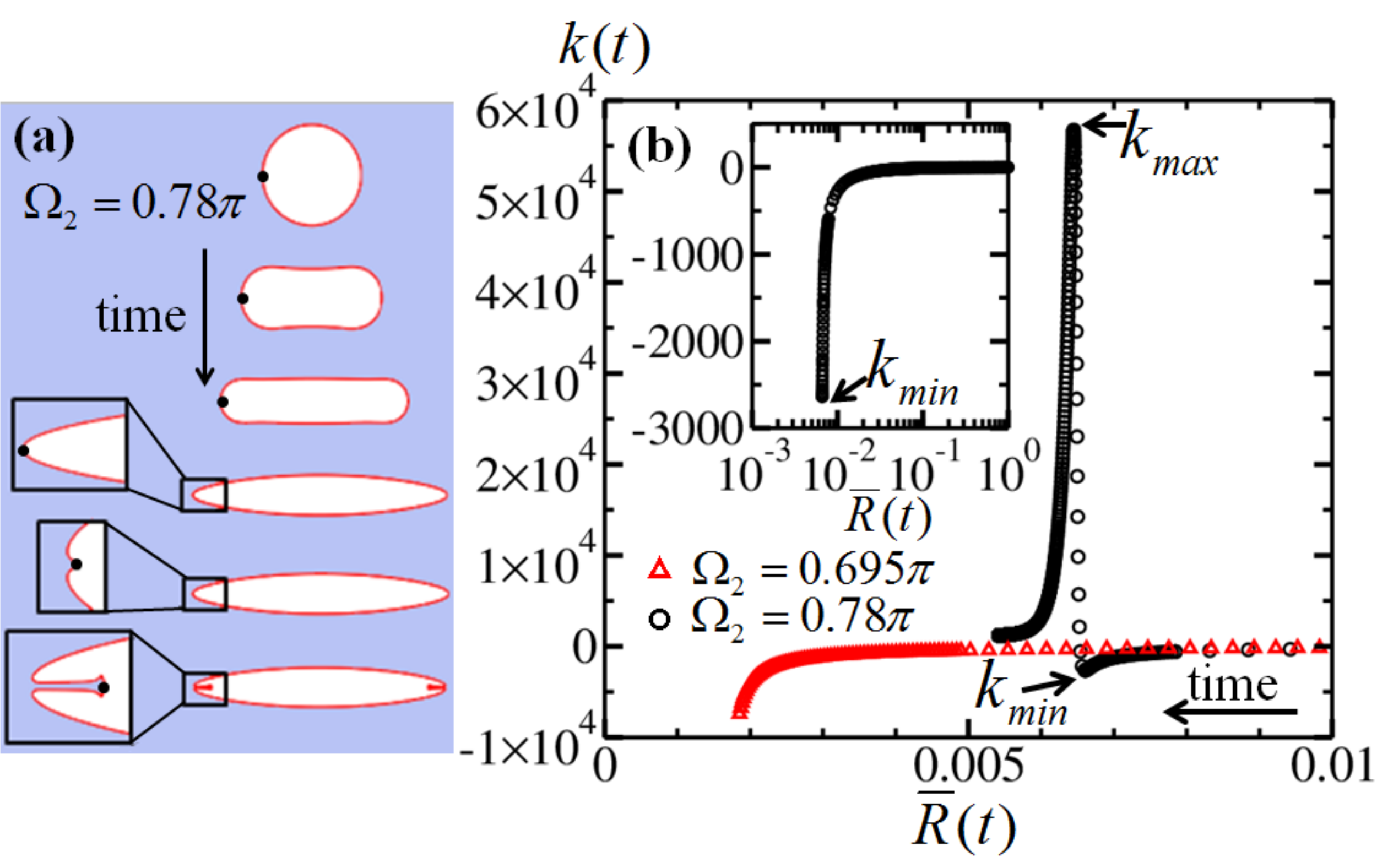}
\caption{\label{fig:kmin_kmax}Cusp-like mode of breakup corresponds to a saddle-node. Here initial amplitude $A_2=0.01$ is fixed. (a) For initial phase $\Omega_2=0.78\pi$, the time evolution of the cross-section shape rescaled by the average radius shows that the interface first evolves towards a shape with sharp tips then deviate from it, consistent with a phase space trajectory controlled by a saddle-node. Sharp tips are followed by re-entrant water fingers. (b) Two types of curvature evolutions for different modes of breakups. The curvature following the west tip (as denoted by black dots in (a), but the result will be the same following the east tip due to the spatial symmetry here) is plotted as a function of the average radius $\bar{R}$. Black open circles: non-monotonic curvature evolution for initial phase $\Omega_2=0.78\pi$. The curvature first reaches a valley, whose absolute value is defined as $k_{min}$. Then it reverses its sign and a re-entrant water finger forms. During the formation of the finger, the tip curvature reaches a peak value $k_{max}$. Red open triangles: for initial phase $\Omega_2=0.695\pi$ where the interface evolves into a double-point coalescence (Fig. \ref{fig:shapes} (c)), the curvature following the west tip of the cross-section decreases monotonically until the coalescence happens. Inset: a closer look at the tip curvature evolution for $\Omega_2=0.78\pi$ until the emergence of $k_{min}$ (in a linear-log plot).}
\end{center}
\end{figure}

In this part, I will first describe the evolution of the interface that forms sharp tips followed by re-entrant water fingers. Starting with single $n=2$ mode perturbation, when the initial amplitude $A_2$ is fixed, for a wide range of initial phases $\Omega_2$, the interface evolves into a coalescence-type breakup as discovered by Turitsyn \emph{et al.} \cite{turitsyn09}. However, for a narrower but finite range of initial phases, as shown in  Fig. \ref{fig:shapes} (d) and Fig. \ref{fig:kmin_kmax} (a), the interface develops sharp tips followed by re-entrant water fingers. In this case, the dynamics is strongly nonlinear. The time evolution of the interface can be divided into two successive parts. In the first part, the cross-section first evolves into a long slit-like shape with two sharp tips. The two tips then sharpen quickly with more and more negative curvatures (by definition, the curvature is negative when the tip curves outwards). In the second part, at one point, those two sharp tips reverse their signs of curvatures rapidly, forming two water intrusions. Those two intrusions then grow and form two water fingers intruding into the air. To quantify the sharpening of the tip, I measured the tip curvature $k(t)$ following the tip point (denoted by black dots in Fig. \ref{fig:kmin_kmax} (a)) that is going to become sharp. The tip curvature $k(t)$ changes non-monotonically with time. The two parts during the interface evolution are divided by the emergence of an extreme negative curvature whose absolute value is $k_{min}$ (Fig. \ref{fig:kmin_kmax} (b)). As shown in the inset of Fig. \ref{fig:kmin_kmax} (b), the tip curvature starts at approximately $-1$, for a slightly perturbed circle. It then becomes more and more negative, corresponding to the time evolution of the interface in which the two end tips sharpen. However, for a given initial condition, this sharpening process reaches a state with the most negative curvature, whose absolute value is defined as $k_{min}$. Until the emergence of $k_{min}$, the two sharp tips have the most negative curvature along the interface given a certain time instance. After the emergence of $k_{min}$, the tip flattens and its curvature increases towards zero. As shown in Fig. \ref{fig:kmin_kmax} (b) (black open circles), eventually at one point, the curvature reverses its sign, which corresponds to the formation of a re-entrant water finger. During the finger's formation, the tip curvature reaches a peak value defined as $k_{max}$ that is much larger than $k_{min}$. After the tip curvature reaches $k_{max}$, the front of the finger starts to broaden, in which case the tip curvature decreases from its peak value. This non-monotonic variation of curvature is also observed for all Lagrangian points in a finite range around the tip.

\begin{figure}
\begin{center}
\includegraphics[width=8cm]{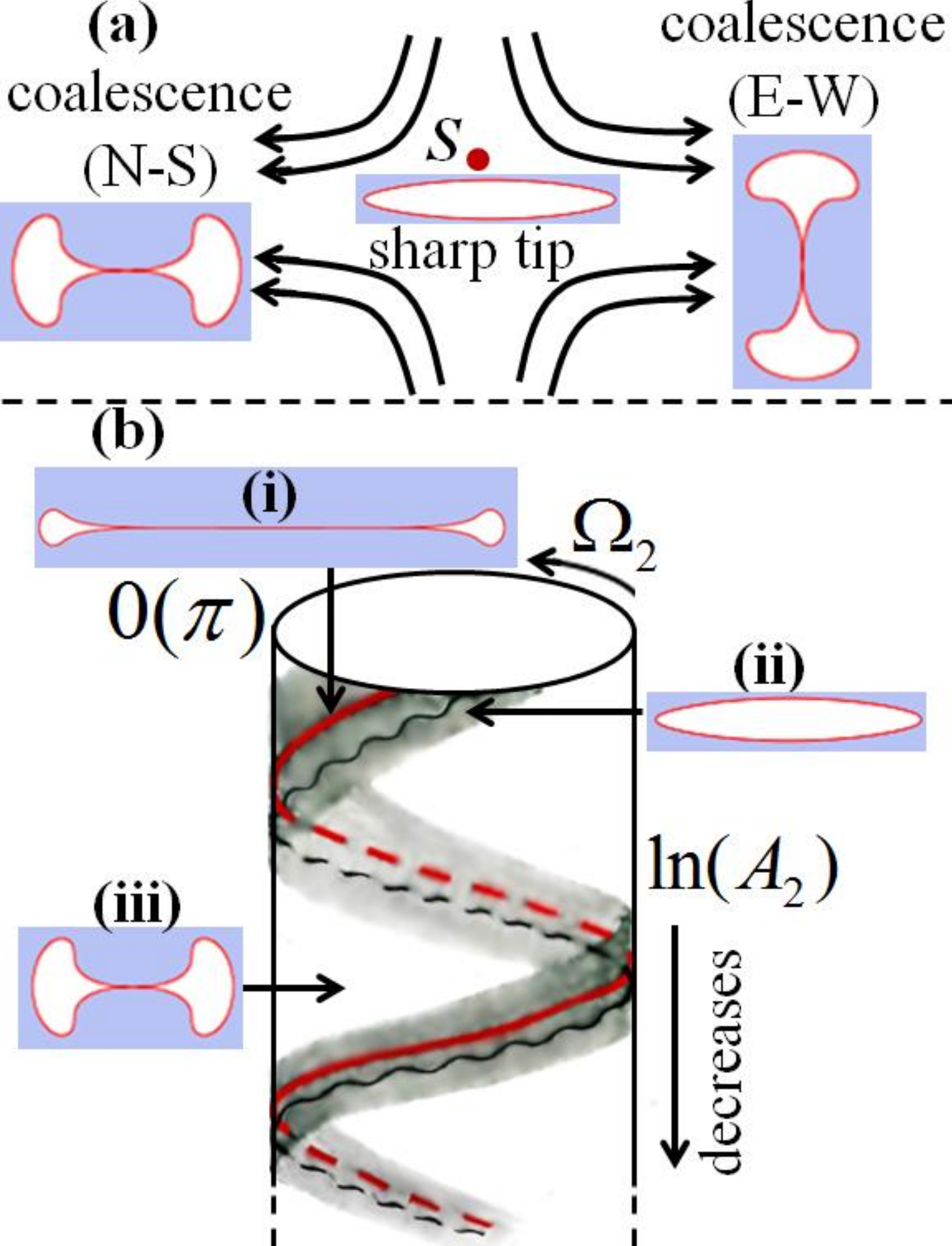}
\caption{\label{fig:schematic}Schematic: (a) saddle-node evolution. A saddle-node $S$ that corresponds to an interface evolution towards a shape with sharp tips separates coalescence-type breakups with distinct contact orientations: coalescence between north and south sides (N-S) and coalescence between east and west sides (E-W). (b) Structure in initial parameter space spanned by $A_2$ and $\Omega_2$. The initial parameter space is represented by the surface of a semi-cylinder. The amplitude $A_2$ decreases moving downwards along the cylinder and the phase $\Omega_2$ winds around the cylinder with period $\pi$. Red curve (the one shape (i) points to): threshold values of initial conditions for which the interface appears to evolve into the curvature singularity. But the singularity is cut off by coalescence (e.g., shape (i)). Black wavy curve: border of the cut-off. To its right the tip curvature varies non-monotonically with time (e.g., black open circles in Fig. \ref{fig:kmin_kmax} (b)). To its left the curvature evolution towards $k_{min}$ is cut off by coalescence and the tip curvature decreases monotonically (red open triangles in Fig. \ref{fig:kmin_kmax} (b)). On the black curve, the maximum value of $k_{min}$ for a given perturbation amplitude $A_2$ is attained (e.g., inset of Fig. \ref{fig:kmin_otheramp}). Gray band: initial conditions for which complex breakups are observed. For some of them, the interface develops sharp tips (e.g., shape (ii)). Outside of the band, the generic outcome is coalescence-type breakups (e.g., shape (iii)).} 
\end{center}
\end{figure}

As described above, such a formation of sharp tips is consistent with a phase space trajectory controlled by a saddle-node, in the sense that the interface evolution towards a shape with sharp tips invariably deviate from it later after an extreme value of curvature $k_{min}$ is reached. This dynamical structure is sketched in Fig. \ref{fig:schematic} (a). Fig. \ref{fig:schematic} (a) also shows that such a cusp-like mode of breakup separates coalescence types of breakups with distinct coalescence orientations. In the case of starting with initial conditions including only an $n=2$ mode perturbation, there are two possible coalescence orientations: coalescence between east and west sides (E-W), and between south and south sides (N-S).

\begin{figure}
\begin{center}
\includegraphics[width=12.5cm]{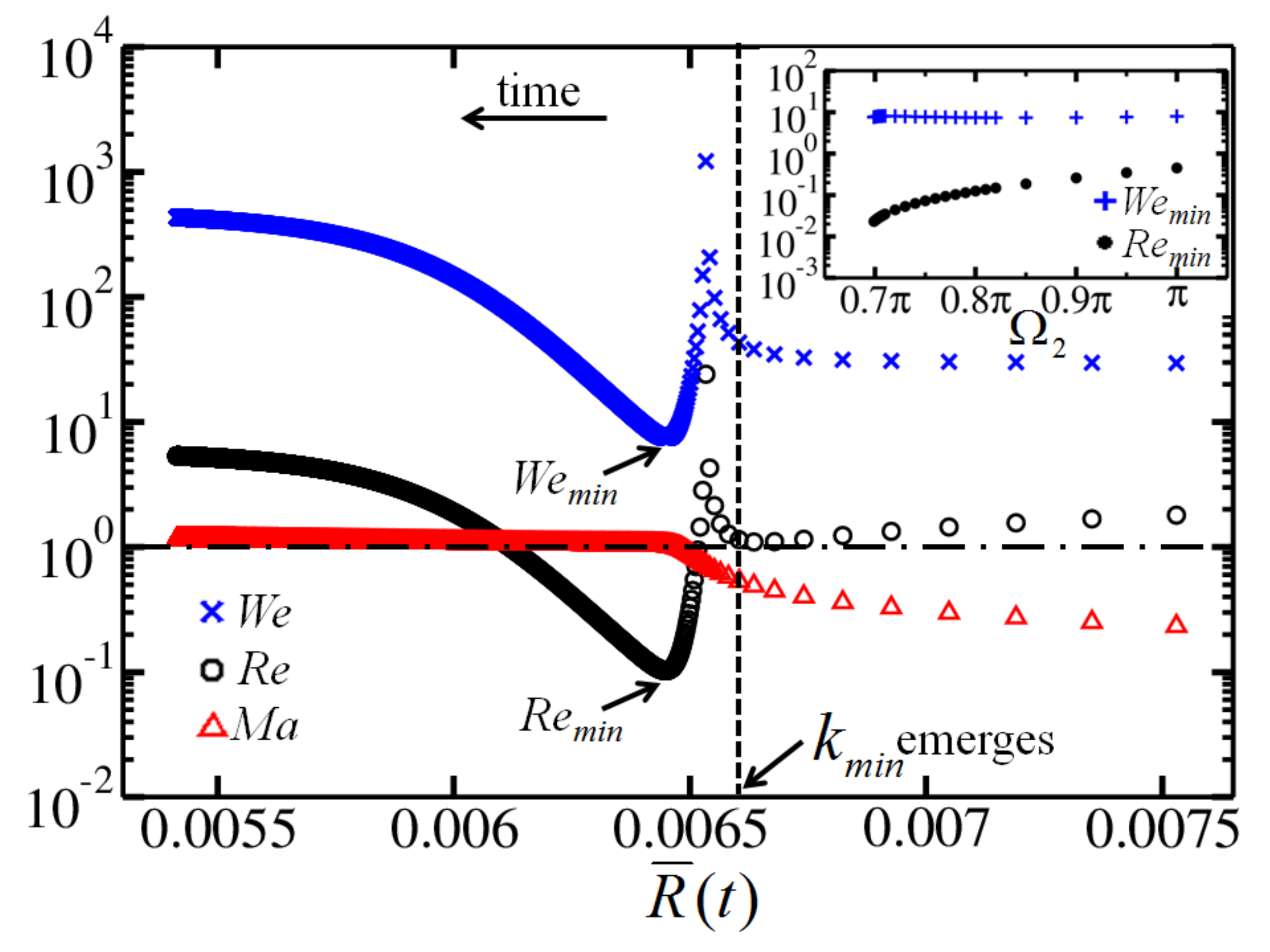}
\caption{\label{fig:re_we} Effects of surface tension, air viscosity and air compressibility before and after the emergence of the extreme tip curvature $k_{min}$. For a given initial condition ($A_2=0.01$, $\Omega_2=0.78\pi$ here), the Weber number $We$ (blue crosses), Reynolds number $Re$ (black open circles) and Mach number $Ma$ (red open triangles) at the tip are plotted as functions of the average radius $\bar{R}(t)$ in a log-linear plot. The average radius for which $k_{min}$ emerges is marked by the vertical dashed line. The horizontal dashed-dotted line indicates the value of unity. After the emergence of $k_{min}$ (to the left of the vertical dashed line), both the $We$ and $Re$ show minima as indicated by $We_{min}$ and $Re_{min}$ respectively. Inset: the minimum Weber number $We_{min}$ (blue pluses) and the minimum Reynolds number $Re_{min}$ (black solid circles) are plotted as functions of the initial phase in a log-linear plot, while the initial amplitude $A_2=0.01$ is fixed.} 
\end{center}
\end{figure}

For the second part -- formation of the re-entrant water finger, Fig. \ref{fig:finger} shows the features of the re-entrant water finger I observed from simulations. Here I use the initial condition $A_2=0.01$ and $\Omega_2=0.78\pi$. For a different initial condition, the resulting water finger shares qualitatively the same features as I show here but with a different width and growth rate. The velocity of water within the finger is approximately one order of magnitude higher than the velocity of water at other parts along the interface. This can be seen in the shape sequence in Fig. \ref{fig:kmin_kmax} (a): after the east and west end tips reach $k_{min}$ (the last three cross-section shapes in Fig. \ref{fig:kmin_kmax} (a)), the only noticeable change along the interface is the formation of the re-entrant water finger. The water finger grows and forms a mushroom-like front around the finger tip (the last cross-section shape in Fig. \ref{fig:kmin_kmax} (a)). The edges of the mushroom-like front possess the maximum curvatures along the interface after they form. These high curvatures exceed the spatial resolution of my current simulations and require an improved simulation with better spatial resolution in future work. On the one hand, the dynamics of the water finger is a very interesting topic itself. On the other, however, due to its high speed and extremely high curvature, if the water finger were to form in the bubble breakup experiments \cite{keim06, schmidt09, keim11}, one should expect it to be more strongly affected by other physical effects that are not included in my current model, such as surface tension, viscous dissipation and the compressibility of air. This concern is further confirmed by looking at the Weber number $We$, the Reynolds number $Re$ and Mach number $Ma$ at the finger tip. The relative effects of inertia with respect to  surface tension and viscosity are described by the Weber number $We$ and  Reynolds number $Re$ respectively. The importance of air compressibility is measured by the Mach number $Ma$. The three numbers are defined as follows:

\begin{figure}
\begin{center}
\includegraphics[width=14cm]{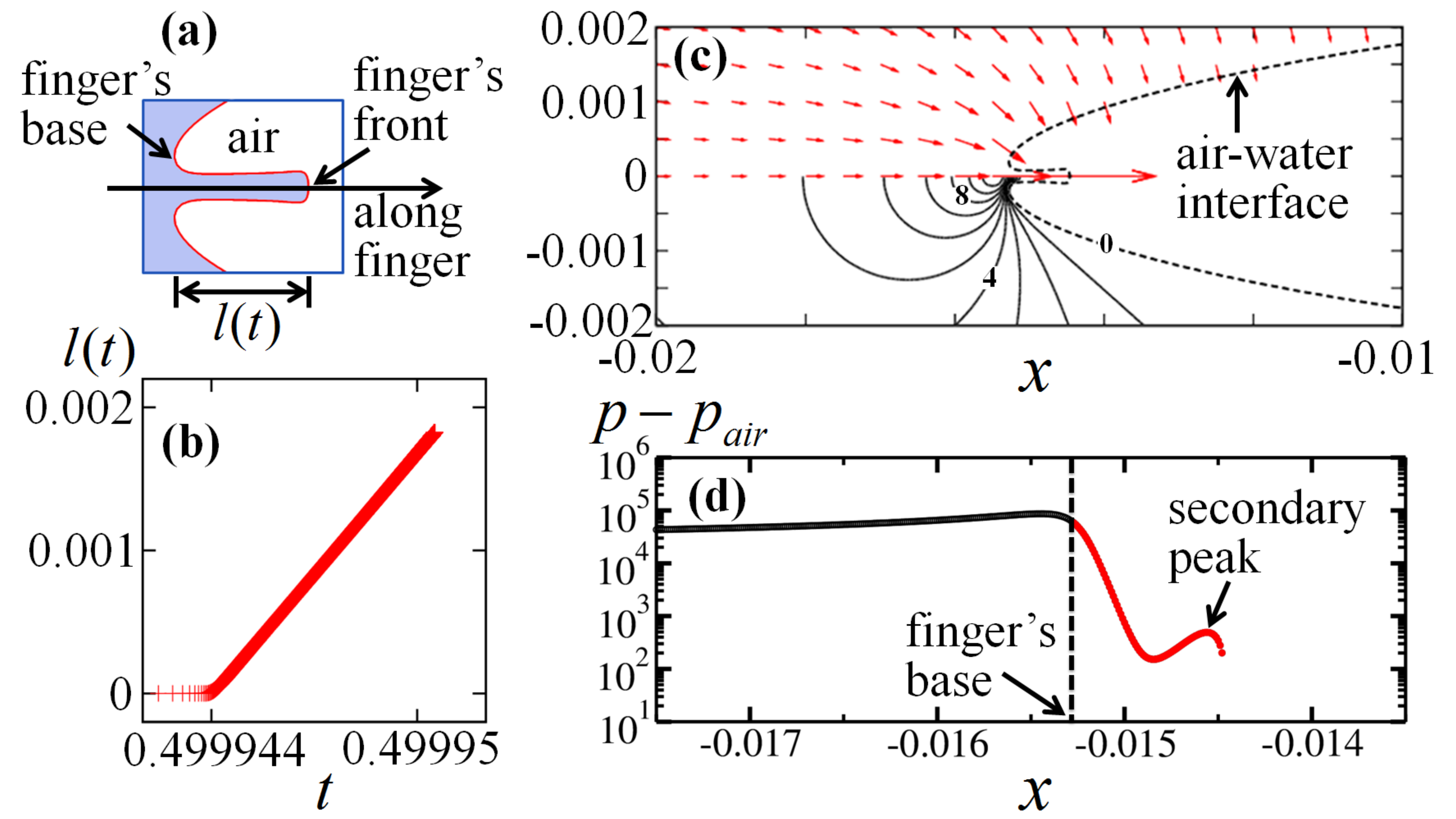}
\caption{\label{fig:finger}A re-entrant water finger forms from the tip after the tip reaches an extreme curvature $k_{min}$. For initial amplitude $A_2=0.01$ and phase $\Omega_2=0.78\pi$, here I show (a) a closer look at the water finger at one time instance during its formation. The blue region represents water. (b) The length of the finger $l(t)$ vs. time. the finger grows approximately linearly in time. (c) The velocity field and pressure contours around the finger taken at the same time as the finger in (a). The dashed line indicates the air-water interface. The upper half of the plot shows the velocity field around and within the finger. The arrow indicates the direction of the velocity ${\bf u}$ at a field point, and the length of the arrow equals to $2\times 10^{-6}|{\bf u}|$. The lower half of plot (c) shows the pressure contours around the finger. The pressure is measured relative to the air pressure $p_{air}$. Adjacent contour lines have an equal spacing that is approximately $2MPa$ in bubble breakup experiments. (d) Pressure along the finger on the symmetry axis at the same time as that in (a) and (c). The vertical dashed line marks the position of the finger's base. The pressure drops significantly from its peak value (just outside of the finger) within the finger (to the right of the dashed line, red curve) and forms a secondary peak behind the front of the finger. The secondary peak is consistent with the formation of a mushroom-like front of the finger at a later time.} 
\end{center}
\end{figure}

\begin{equation}
We=\frac{\rho U^2 L}{\sigma},\quad Re=\frac{U L}{\nu},\text{ }and\text{ }Ma=\frac{U}{C_g}
\label{eqn:re_we_ma}
\end{equation} 
Here, $\rho$ is the density of water. $U$ and $L$ are the characteristic velocity and length scales in the problem respectively. $\nu$ is the kinematic viscosity of air or water. $\sigma$ is the surface tension, and $C_g$ is the sound speed in air. In my case here, $U$ equals the speed of the tip point, and $L$ equals the radius of curvature at the tip. For other parameters, I use $\rho=1000kg/m^3$, $\sigma=72.8\times10^{-3}N/m$ and $C_g=340m/s$ for experiments at room temperature. Since the kinematic viscosity of air is about $15$ times higher than that of water, the viscous dissipation in air is expected to have a more significant effect than in water, and hence here I only consider the viscosity of air and set $\nu=\nu_{air}=15\times10^{-6}m^2/s$. To calculate the three numbers as plotted in Fig. \ref{fig:re_we}, all the non-dimensionalized quantities obtained from simulations are converted to dimensional quantities using scales in experiments as estimated in Section \ref{section:problem}. The effects of surface tension, air viscosity and compressibility will be significant when the corresponding numbers $We$, $Re$ and $Ma$ reach unity.

The formation of the re-entrant water finger corresponds to the region to the left of the vertical dashed line in Fig. \ref{fig:re_we}. It shows that, along the $We$, $Re$ vs. $\bar{R}$ curves, they both show spikes immediately after the emergence of $k_{min}$. This corresponds to where the curvature reverses its sign and passes zero. After that, each curve ($We$, $Re$ vs. $\bar{R}$) show a minimum. For the Weber number $We$, this minimum is still relatively large compared with that of the Reynolds number $Re$ and this minimum of $We$ remains approximately constant for different initial phases (inset of Fig. \ref{fig:re_we}). Thus, during the finger formation, the effect of surface tension remains relatively small, but both the air viscosity ($Re$) and compressibility ($Ma$) will be important, as Fig. \ref{fig:re_we} suggests. On the other hand, all these complexities during the finger formation arise after the emergence of $k_{min}$. If $k_{min}$ diverges, the dynamics afterwards is expected to be changed. For the two reasons above, I will focus on the curvature singularity suggested by the divergence of $k_{min}$ in the rest of the paper.

\subsection{\label{subsection:curvature_phasespace}Curvature singularity: phase space behavior}
In this part, I focus on the phase space behavior for different initial conditions corresponding to the interface evolution into a near singular shape. What I found is that the interface evolution is controlled by a saddle-node that corresponds to a finite-time curvature singularity. In addition, I will show that this curvature singularity is pre-empted by coalescence and can only be realized with vanishingly small perturbation amplitude.

\begin{figure}
\includegraphics[width=12.5cm]{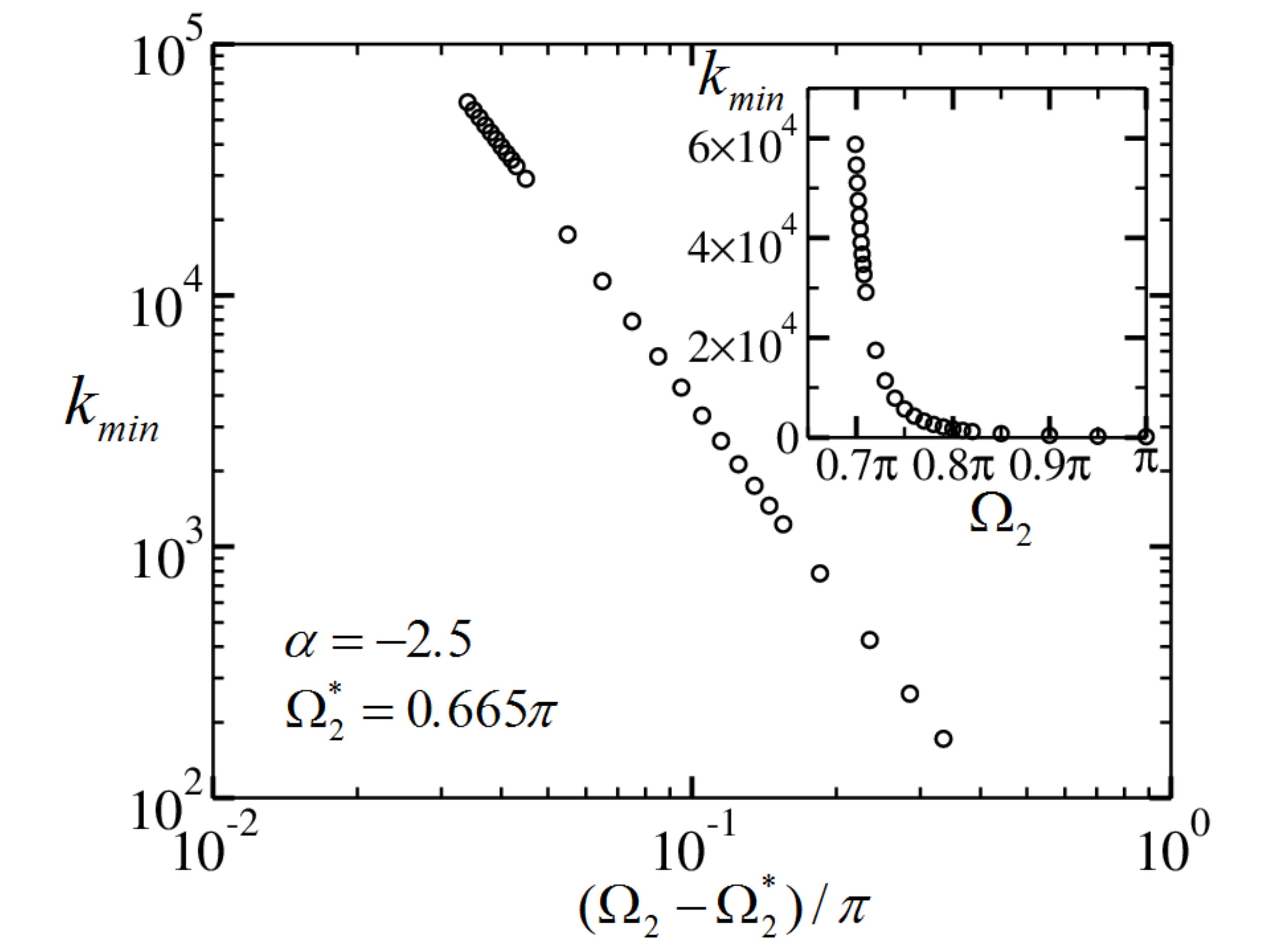}
\caption{\label{fig:divergence}Divergence of extreme tip curvature $k_{min}$ defined in Fig. \ref{fig:kmin_kmax} (b) as a function of initial phase $\Omega_2$. A log-log plot for $k_{min}$ vs. ($\Omega_2-\Omega_2^*$)$/\pi$ shows that $k_{min}$ appears to diverge at a threshold value $\Omega_2^*=0.665\pi$ of the initial phase. The divergence of $k_{min}$ is characterized by an exponent $\alpha=-2.5$. Inset: the divergence of $k_{min}$ as a function of initial phase $\Omega_2$ (plotted on a linear scale).}
\end{figure}

I start by focusing on a subset of initial conditions, in which I fix the initial amplitude $A_2=0.01$ and only vary the initial phase $\Omega_2$. The emergence of $k_{min}$ marks a turning point during the curvature evolution. The value of $k_{min}$ appears to diverge as the initial phase $\Omega_2$ decreases towards around $0.7\pi$ (inset of Fig. \ref{fig:divergence}). To quantify this divergence, I fit $k_{min}$ into a power function of the initial phase as  $k_{min}\sim(\Omega_2-\Omega_{2}^*)^{\alpha}$. Within the data range, as shown in Fig. \ref{fig:divergence}, I found that the divergence of $k_{min}$ is described by a threshold value $\Omega_2^*=0.665\pi$, and the corresponding scaling exponent is $\alpha=-2.5$ whose origin still requires further investigations. Thus, on the one hand, around the emergence of such an extreme curvature $k_{min}$, the tip evolution that first sharpens then broadens (Fig. \ref{fig:kmin_kmax} (b) (black open circles)) is consistent with a phase space trajectory controlled by a saddle point. On the other, as I tune the initial condition (in this case, fix the amplitude and reduce the phase), the value of $k_{min}$ appears to diverge. This suggests that the saddle point corresponds to a curvature singularity. By tuning the initial condition towards threshold values, I expect to attain phase space trajectories closer to the curvature singularity. The threshold value $\Omega_2^*=0.665\pi$ indicates that if I start with the initial phase at this value, the interface is expected to develop infinitely sharp tips. 

\begin{figure}
\begin{center}
\includegraphics[width=14cm]{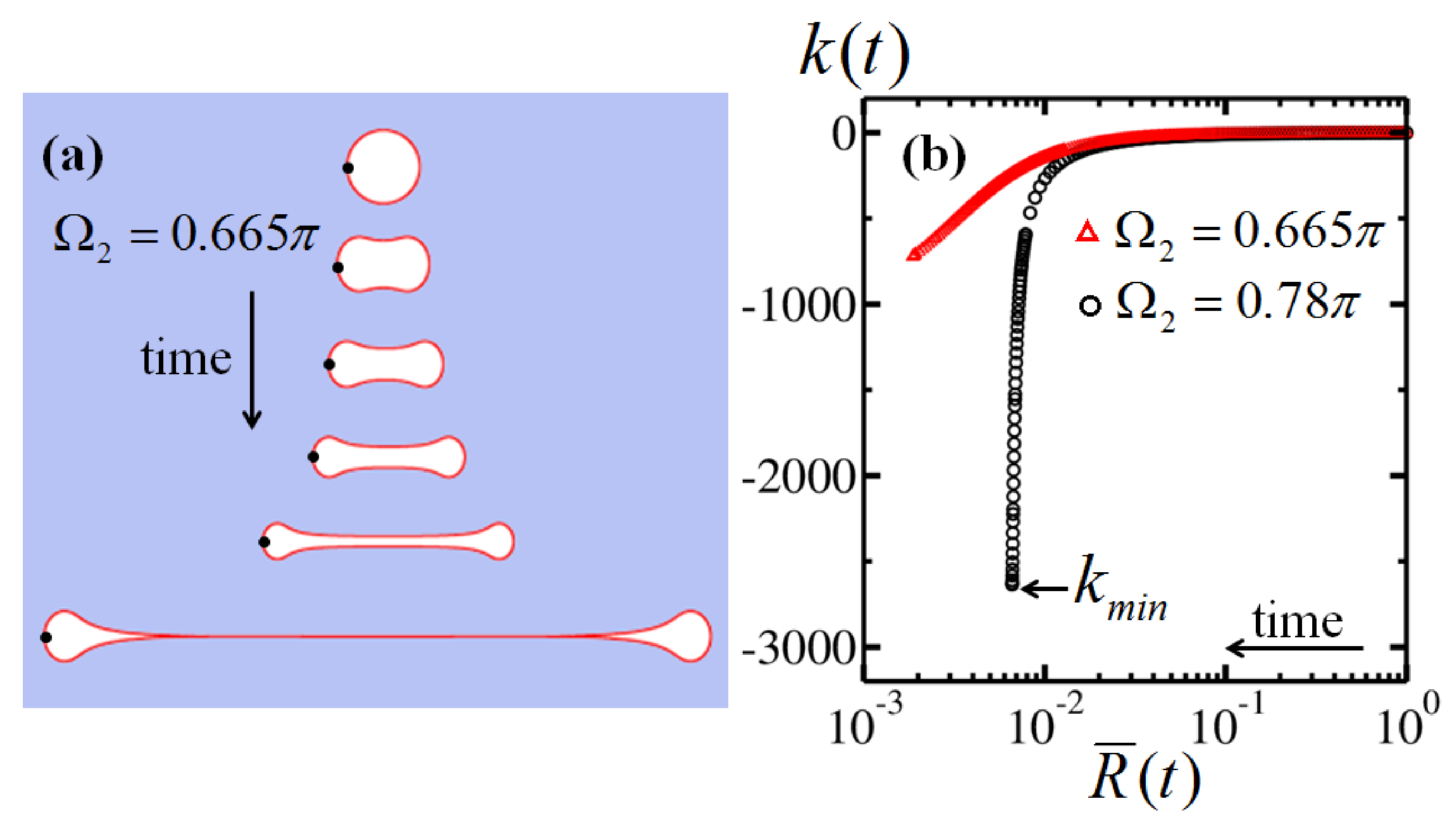}
\caption{\label{fig:cutoff}Interface evolution towards the curvature singularity is pre-empted by coalescence. (a) Starting with the initial phase at the threshold value, the time evolution of the cross-section shape rescaled by the average radius for initial amplitude $A_2=0.01$ and phase $\Omega_2=\Omega_2^*=0.665\pi$ shows that the interface evolves into a coalescence before the curvature at the end tip (denoted by black dots) diverges. (b) For this initial condition, the curvature of the tip (denoted by black dots in (a)) only reaches a relatively small value (red triangles) and its divergence is cut off by coalescence. On the contrary, for the initial condition $A_2=0.01$ and $\Omega_2=0.78\pi$, the tip curvature reaches an extreme value $k_{min}$ (black circles) before it reverses its sign. The value of $k_{min}$ attained here is about $4$ times larger than the most negative curvature attained for $\Omega_2=0.665\pi$ (red triangles).}
\end{center}
\end{figure}

However, when starting with initial condition at the threshold value, the interface actually evolves into a coalescence before the tip curvature diverges. In this case, the time evolution of the cross-section shape is plotted in Fig. \ref{fig:cutoff} (a) showing that the cross-section gets elongated but the divergence of the tip curvature is cut off by a coalescence between the north and south sides of the interface. For cross-section shapes shown in Fig. \ref{fig:cutoff} (a), the time evolution of the tip curvature is plotted in Fig. \ref{fig:cutoff} (b) (red triangles) together with the tip curvature evolution for initial phase $\Omega_2=0.78\pi$ (black circles, and I only show the first part of the evolution until the emergence of $k_{min}$). In the latter case ($\Omega_2=0.78\pi$) the curvature evolution is not pre-empted by coalescence. Fig. \ref{fig:cutoff} (b) shows that for initial phase at the threshold value, the tip curvature decreases monotonically and the most negative value it can reach before coalescence is much smaller in absolute value than that for $\Omega_2=0.78\pi$, which is not even very close to the threshold value. Hence I have shown here that the time evolution towards such a curvature singularity is pre-empted by coalescence and the singularity cannot be realized. In addition, for a finite range of initial phases ($0.665\pi < \Omega_2 < 0.699\pi$), the decrease of tip curvature towards the turning point $k_{min}$ is cut off by coalescence. Thus, as the initial phase decreases towards $0.665\pi$, the maximum value of $k_{min}$ is obtained for $\Omega_2=0.699\pi$ corresponding to the left-most point in Fig. \ref{fig:divergence}.   

\begin{figure}
\includegraphics[width=12.5cm]{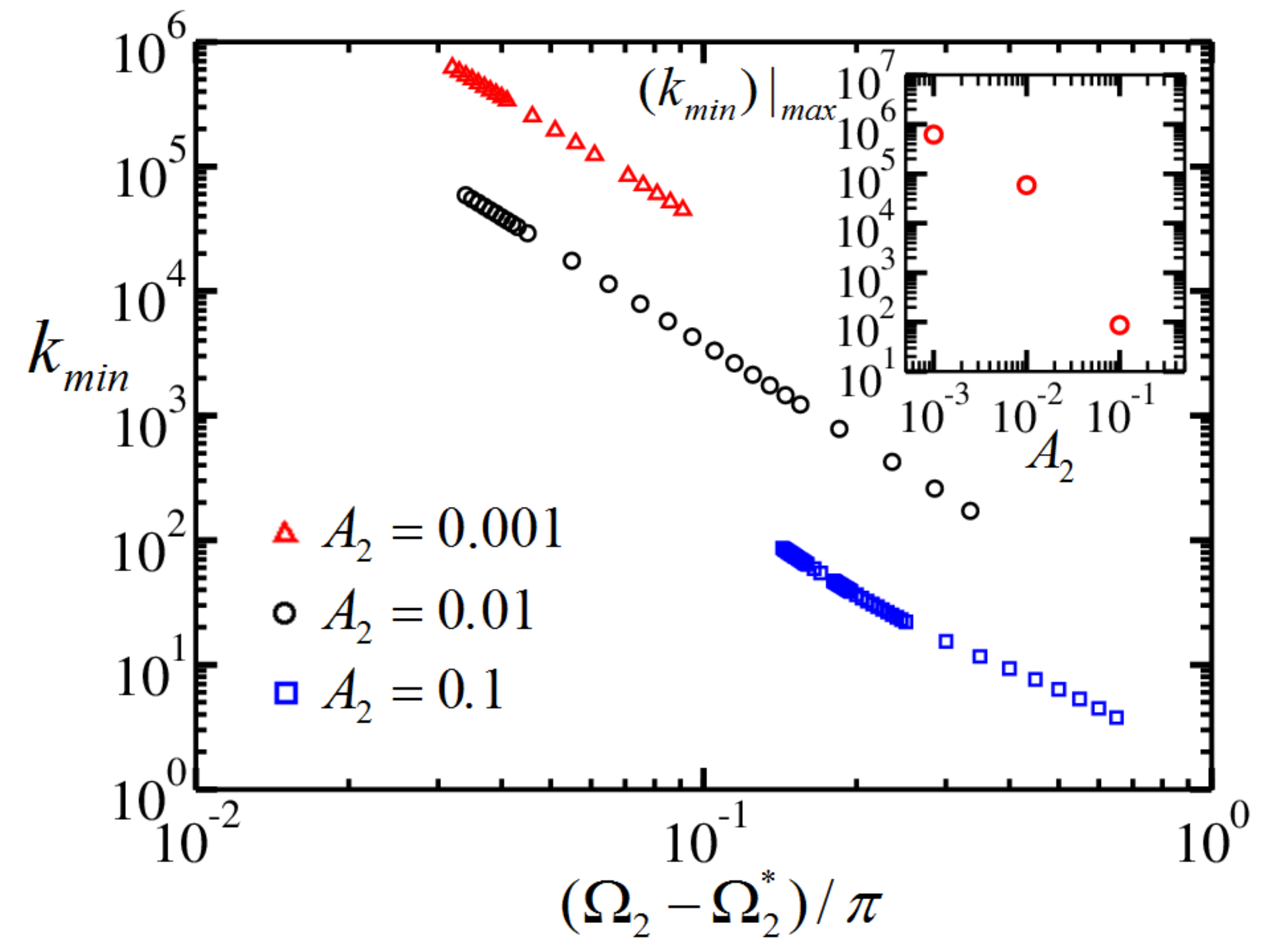}
\caption{\label{fig:kmin_otheramp}Divergences of the extreme tip curvature $k_{min}$ as the initial phase varies for three different initial perturbation amplitudes: $A_2=0.001$ (red open triangles, top curve), $0.01$ (black open circles, middle curve) and $0.1$ (blue open squares, bottom curve). Smaller perturbation amplitude leads to larger maximum value of $k_{min}$. The exponent $\alpha=-2.5$ obtained with initial perturbation amplitude $A_2=0.01$ is used to determine the threshold value $\Omega_2^*$ for the other two perturbation amplitudes $A_2=0.001$ and $A_2=0.1$. This results in a threshold value $\Omega_2^*=0.934\pi$ for $A_2=0.001$, and $\Omega_2^*=0.3\pi$ for $A_2=0.1$. For all three amplitudes, starting with the initial phase at the threshold value, the interface evolution towards the curvature singularity is pre-empted by coalescence. Inset: the maximum value of the extreme tip curvature $k_{min}$ (denoted as $(k_{min})|_{max}$) obtained for a fixed initial amplitude (e.g., the left-most point on each curve in the main figure) as a function of the initial perturbation amplitude $A_2$.}
\end{figure}

So far, for single $n=2$ mode perturbation, I have only shown the dynamics for a subset of initial conditions by just varying the phase. For the amplitude $A_2=0.01$, the curvature singularity is pre-empted by coalescence. However, if I am also allowed to vary the amplitude, the question is, in the initial parameter space spanned by $A_2$ and $\Omega_2$, whether there exists a triple point at which the curvature singularity is realized. My numerical results suggest that such a triple point does not exist for finite-size perturbations. The same dynamics as the one for $A_2=0.01$ is observed for perturbation amplitudes fixed at different values: when the initial amplitude is fixed, for a range of initial phases, the interface develops sharp tips followed by re-entrant water fingers. The tip curvature evolution is qualitatively the same as demonstrated in Fig. \ref{fig:kmin_kmax} (b) (black open circles). Here I show the divergence of the extreme tip curvature $k_{min}$ for two different initial amplitudes, $A_2=0.001$ and $A_2=0.1$. Both of them show that $k_{min}$ diverges as the initial phase decreases. This divergence is then quantified by fitting $k_{min}$ into a power function of $(\Omega_2-\Omega_2^*)$ with exponent $\alpha$. In the fitting, I use the same exponent $\alpha=-2.5$ obtained in previous analysis for $A_2=0.01$ to determine the threshold value $\Omega_2^*$ of the initial phase. This gives me a threshold value $\Omega_2^*=0.934\pi$ for $A_2=0.001$ and $\Omega_2^*=0.3\pi$ for $A_2=0.1$. The threshold values $\Omega_2^*$ for the three different amplitudes ($A_2=0.001$, $0.01$ and $0.1$) are then used to plot $k_{min}$ as a function of $(\Omega_2-\Omega_2^*)$ in Fig. \ref{fig:kmin_otheramp}. At each threshold value, the interface appears to evolve into a curvature singularity. But similar to the dynamics for $A_2=0.01$, for either $A_2=0.001$ or $0.1$, when starting with initial phase at the threshold value, the interface evolves into a coalescence before the tip curvature diverges and the curvature singularity cannot be realized. This cut-off due to coalescence is again observed in a finite range of initial phases as the phase decreases towards the threshold value. The cut-off happens for $0.934\pi<\Omega_2<0.966\pi$ for $A_2=0.001$, and for $0.3\pi<\Omega_2<0.443\pi$ for $A_2=0.1$ respectively. Even though all the three curves agree with an exponent $\alpha=-2.5$ in characterizing the divergence of $k_{min}$ as the initial phase approaches the threshold value, it is noticed that for $A_2=0.1$, the data that satisfy this power-law scaling are only within a small dynamic range (the left portion of the blue squares in Fig. \ref{fig:kmin_otheramp}). One possible explanation is that the interface evolution for larger perturbation amplitudes ($A_2=0.1$ here) is further away from the curvature singularity, with a much smaller $k_{min}$ compared with that for smaller amplitudes (e.g., $A_2=0.001$ and $0.01$). However, extending the dynamic range closer to the threshold value is limited by the fact that the evolution of tip curvature towards $k_{min}$ is cut off by coalescence.

For different initial amplitudes, there are threshold values of the initial phase for which the interface is expected to develop a sharp tip with infinite curvature. This suggests that in the initial parameter space spanned by $A_2$ and $\Omega_2$, the threshold values as combinations of $A_2$ and $\Omega_2$ fall onto a continuous curve. 
However, numerical results show that starting with a threshold value on this curve, the interface evolves into a coalescence before the tip curvature becomes singular and the curvature singularity cannot be realized. This cut off happens in a finite range around the threshold value. Thus a border can be drawn besides the curve of threshold values. On the border the maximum value of $k_{min}$ is obtained for a given initial perturbation amplitude (the left-most point on each curve in Fig. \ref{fig:kmin_otheramp}). The maximum value of $k_{min}$ for different amplitudes is plotted in the inset of Fig. \ref{fig:kmin_otheramp}. Although the curvature singularity is not realized for the three amplitudes, Fig. \ref{fig:kmin_otheramp} shows that the maximum value of $k_{min}$ attained for different initial amplitudes is higher when the amplitude is smaller, which suggests that the dynamics with smaller initial perturbation amplitude is closer to the curvature singularity when the initial phase is also tuned to an appropriate value. Combined with the numerical result that this curvature singularity is pre-empted by coalescence for finite perturbation amplitudes, this implies that the curvature singularity can only be reached in the limit that the perturbation amplitude goes to zero, i.e., $A_2\to 0$.    

As a short summary here, Fig. \ref{fig:schematic} (b) sketches the structure in the initial parameter space spanned by $A_2$ and $\Omega_2$ when the initial condition only contains one single $n=2$ mode perturbation. Because the structure is periodic along the direction of the initial phase $\Omega_2$ with period $\Delta\Omega_2=\pi$ (here I do not distinguish between the two interface evolutions attained for two initial phases different by $\pi$ because they are identical only up to a rotation by $90^\circ$.) and the amplitude $A_2$ is bounded from above, the initial parameter space is represented as the surface of a semi-cylinder. The initial amplitude decreases as I move downwards along the cylinder and the phase winds around the cylinder with period $\pi$. 

The red curve (the one shape (i)  in Fig. \ref{fig:schematic} (b) points to) consists of threshold values for initial conditions for which the interface appears to evolve into a curvature singularity. However, when starting with initial conditions on the red curve, the curvature singularity is cut off by coalescence (e.g., cross-section shape (i) in Fig. \ref{fig:schematic} (b)) and cannot be realized. The divergence of tip curvature is pre-empted by coalescence not only on the red curve, but also for a continuous band of initial conditions. This is represented by the black wavy curve on the cylinder. To the right of the black curve, the tip first reaches its extreme curvature $k_{min}$ (e.g., cross-section shape (ii) in Fig. \ref{fig:schematic} (b)) and then reverses the sign of curvature, forming a re-entrant water finger. To the left of the black curve, the tip curvature decreases monotonically until it is cut off by coalescence. On the black curve, a maximum value of $k_{min}$ is obtained for each given perturbation amplitude $A_2$ (each horizontal slice of the cylinder). For initial conditions within the gray band, complex outcomes of breakups are observed including the multiple-points coalescence (e.g., Fig. \ref{fig:shapes} (c) (e)) and interface evolution towards sharp tips (e.g., cross-section shape (ii) in Fig. \ref{fig:schematic} (b)). For a wide range of initial conditions outside of the gray band, the interface ends in a coalescence-type breakup (e.g., cross-section shape (iii) in Fig. \ref{fig:schematic} (b)). Numerical results further suggest that the formation of sharp tips (e.g., cross-section shape (ii) in Fig. \ref{fig:schematic} (b)) can be interpreted as a weakly first-order transition which becomes second-order, corresponding to the formation of a finite-time curvature singularity, in the limit that the initial perturbation amplitude approaches zero, the limit as I go down the cylinder towards the negative infinity.

\subsection{\label{subsection:curvature_dynamics}Curvature singularity: dynamics}
Even though the curvature singularity cannot be realized for finite-size perturbations, for each given amplitude, one can still pick out a phase space trajectory as close as possible to the curvature singularity (corresponding to the maximum value of $k_{min}$) and study the divergences of relevant quantities as the singularity is approached.

\begin{figure}
\includegraphics[width=14cm]{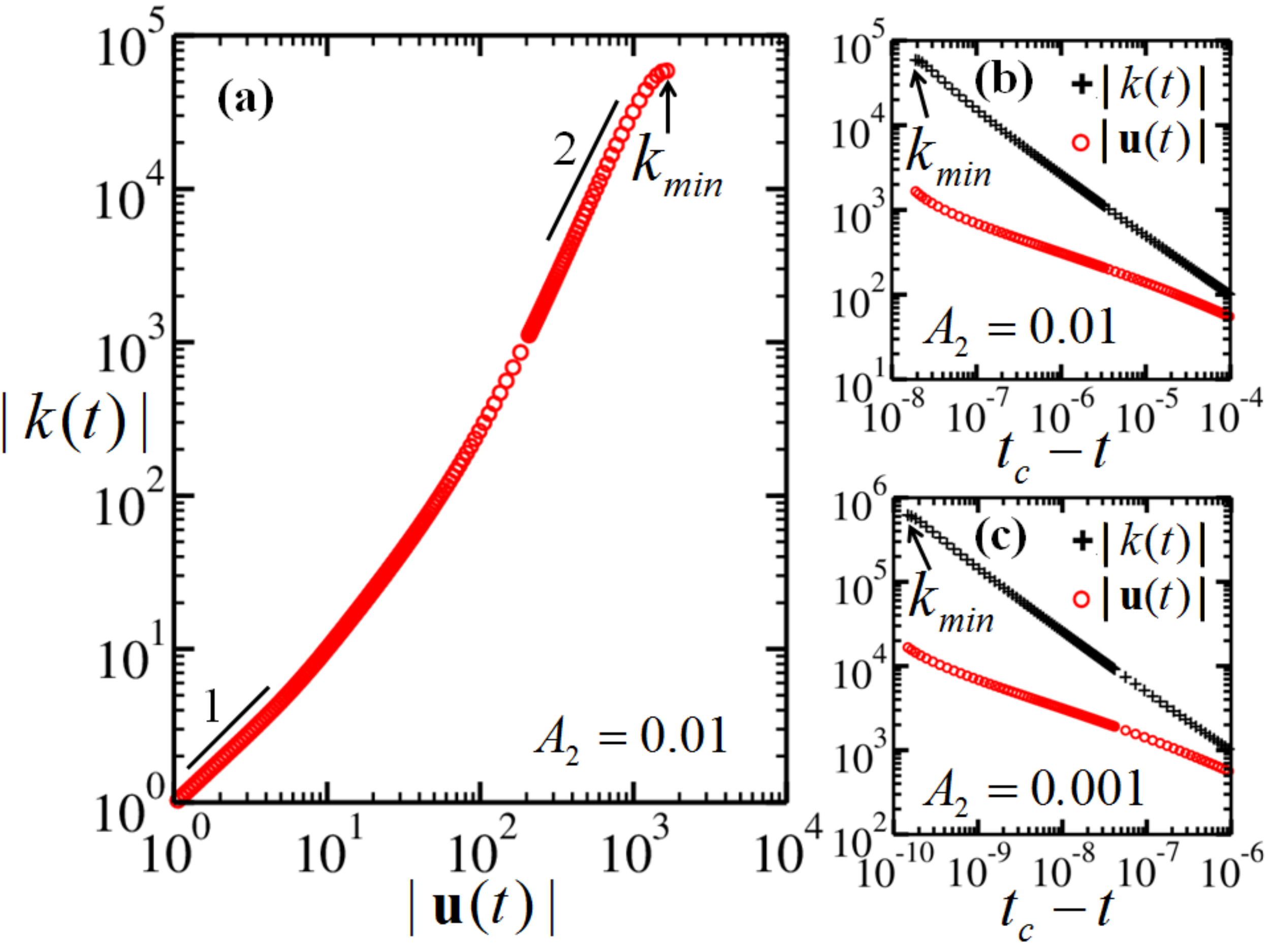}
\caption{\label{fig:k_v}Scalings of the tip curvature and the tip velocity as the curvature singularity is approached. (a) Tip curvature $|k(t)|$ vs. tip velocity $|{\bf{u}}(t)|$ on log scales. I track the evolution until the emergence of $k_{min}$. The initial amplitude is $A_2=0.01$ and the initial phase is $\Omega_2=0.699\pi$ (corresponding to the maximum $k_{min}$ for $A_2=0.01$). To guide the eye, two line segments besides the curve have slope $1$ (bottom-left) and $2$ (upper-right) respectively. (b) For the same initial condition as (a), a plot shows the scalings of $|k(t)|$ and $|{\bf{u}}(t)|$ as functions of $(t_c-t)$, where $t$ is the non-dimensionalized time and $t_c$ corresponds to the onset of the singularity. As the singularity is approached, $|k(t)|$ diverges approximately as $(t_c-t)^{-0.8}$, and $|{\bf{u}}(t)|$ diverges approximately as $(t_c-t)^{-0.4}$. As one may expect, the last a few points close to the emergence of $k_{min}$ where the curvature divergence slows down deviate from the scalings mentioned here. (c) For initial condition $A_2=0.001$ and $\Omega_2=0.966\pi$, the divergences of $|k(t)|$ and $|{\bf{u}}(t)|$ agree with the same exponents found earlier, i.e., $-0.8$ and $-0.4$ respectively.} 
\end{figure}

Fig. \ref{fig:k_v} traces the divergences of the tip curvature $k$ and the speed $|{\bf u}|$ of the tip point as functions of time $t$ until the emergence of $k_{min}$. Here I am interested in the initial phase $\Omega_2$ where the maximum $k_{min}$ is attained for a given initial perturbation amplitude $A_2$. For initial condition $A_2=0.01$ and $\Omega_2=0.699\pi$, when the tip curvature $k$ is plotted against the tip speed $|{\bf u}|$ in a log-log plot, in Fig. \ref{fig:k_v} (a), it shows that the divergences of $k$ and $|{\bf u}|$ go through two different power-law scaling regions before the emergence of $k_{min}$. In the initial moment, when both $k$ and $|{\bf u}|$ are small, their increases are dominated by the average collapse because the perturbation size is still small compared with the average size $\bar{R}$ of the cross-section. In this case, both $k$ and $|{\bf u}|$ scale as $\bar{R}^{-1}$. Hence the slope of the curve in Fig. \ref{fig:k_v} (a) is approximately $1$ initially (bottom-left portion of the curve). However, as the curvature singularity is approached, the nonlinear effect becomes strong, and the $|k|$ vs. $|{\bf u}|$ curve changes its slope from $1$ to approximately $2$, suggesting that the tip curvature $k$ diverges as the square of the tip speed $|{\bf u}|$. I will focus on the  scalings of $k$ and $|{\bf u}|$ while the singularity is approached, i.e., the part of the curve in Fig. \ref{fig:k_v} (a) where the slope is $2$. However, it should be noticed that the dynamic range of the data in this part is limited by the initial evolution dominated by the average collapse ($k$, $|{\bf u}|\sim \bar{R}^{-1}$), and the emergence of $k_{min}$. The scalings I reported below are understood as approximations reflecting this limitation. Using the data that correspond to the portion of the curve with slope $2$ in Fig. \ref{fig:k_v} (a), I first fit $k$ into a power function of $(t_c-t)$, where $t$ is the non-dimensionalized time and $t_c$ corresponds to the onset of the curvature singularity. This shows that as the singularity is approached, the tip curvature $k$ diverges with an exponent approximately $-0.8$, i.e., $k\sim (t_c-t)^{-0.8}$ (Fig. \ref{fig:k_v} (b)). Using the same value of $t_c$, the exponent that describes the divergence of the tip speed $|{\bf u}|$ is obtained, approximately $-0.4$ in the case here (Fig. \ref{fig:k_v} (b)). This result agrees with previous observation that the curvature $k$ diverges as the square of $|{\bf u}|$ (Fig. \ref{fig:k_v} (a)). It is noted that, as one may expect, the last a few points close to the emergence of $k_{min}$ where the curvature divergence slows down deviate from the power-law scalings mentioned above. Using the same method, I also looked at the scalings of $k$ and $|{\bf u}|$ for a different initial condition that leads to a phase space trajectory even closer to the singularity. For initial amplitude $A_2=0.001$ and initial phase $\Omega_2=0.966\pi$ (corresponding to the maximum $k_{min}$ for this amplitude), Fig. \ref{fig:k_v} (c) shows that simulation data still agree with the two scaling exponents, $-0.8$ for $k$ and $-0.4$ for $|{\bf u}|$ respectively. Again, the dynamic range of data to obtain such power-law scalings here is limited by the initial divergence dominated by average collapse and the emergence of $k_{min}$. 

The scalings in Fig. \ref{fig:k_v} provide a way to estimate other physical effects that are ignored in my current model, such as surface tension, viscosity and the compressibility of air, if the curvature singularity were to form in experiments. This will be discussed in the next section (Section \ref{section:discussion}).

\begin{figure}
\begin{center}
\includegraphics[width=13.5cm]{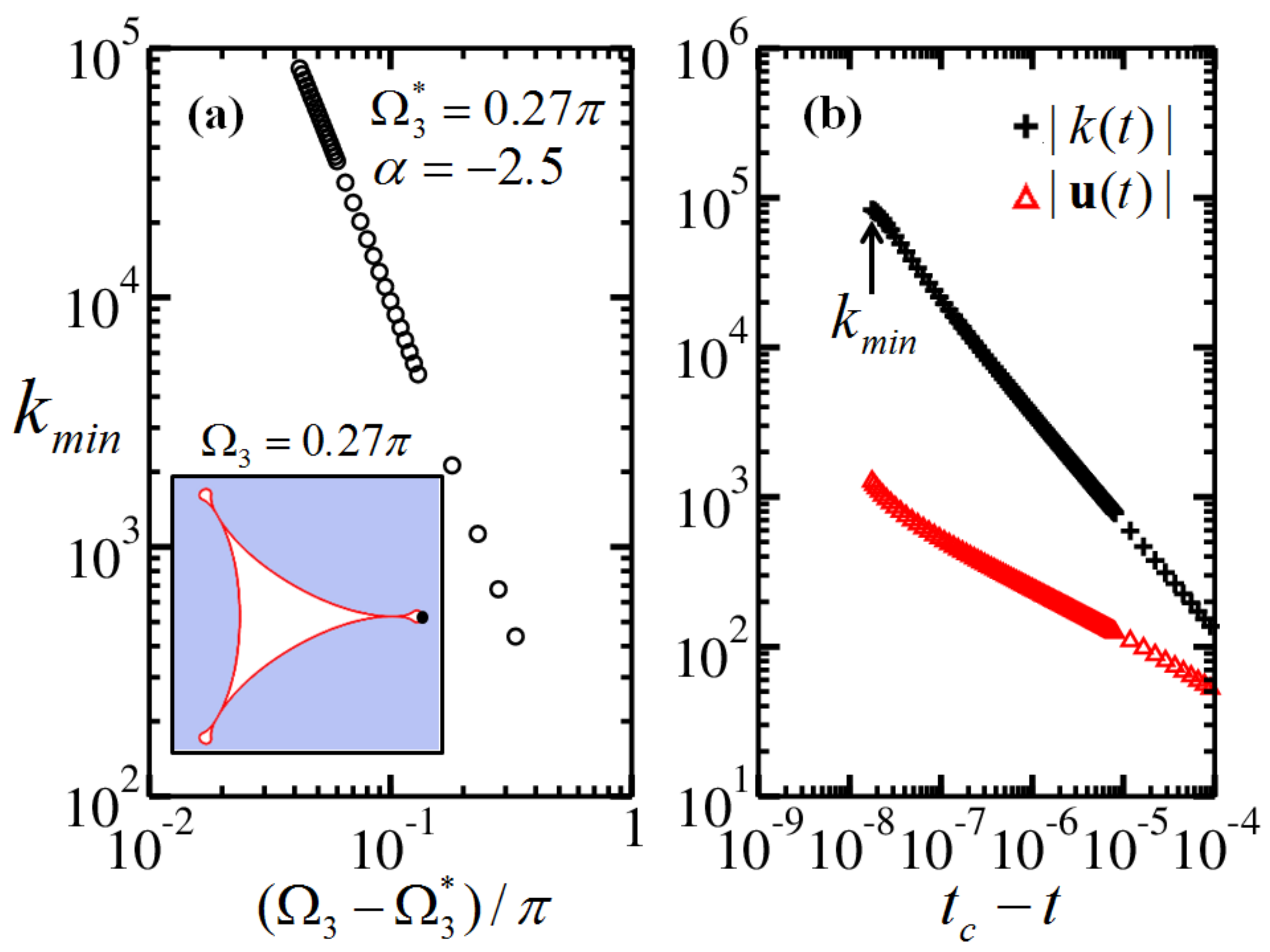}
\caption{\label{fig:n3}Similar dynamics around the curvature singularity is observed for initial perturbations including only one $n=3$ mode. Here I fix the initial perturbation amplitude $A_3=0.01$. (a) Divergence of $k_{min}$, the extreme tip curvature (following the east tip as denoted by the black dot in the inset), as a function of the initial phase $\Omega_3$.  I use the same exponent $\alpha=-2.5$ obtained in the case of starting with an $n=2$ mode perturbation to determine the threshold value $\Omega_3^*$ such that $k_{min}\sim (\Omega_3-\Omega_3^*)^\alpha$. This gives me $\Omega_3^*=0.27\pi$. Inset: starting with initial phase $\Omega_3$ at the threshold value $\Omega_3^*$, the curvature singularity is cut off by coalescence. (b) For initial phase $\Omega_3=0.312\pi$ (corresponding to the maximum $k_{min}$ for $A_3=0.01$), as the curvature singularity is approached, the divergences of the tip curvature $|k(t)|$ and the tip speed $|{\bf u}(t)|$ as functions of $t_c-t$, the time left until the onset of the curvature singularity, agree with the two exponents found earlier, i.e., approximately $-0.8$ and $-0.4$ respectively.}
\end{center}
\end{figure}

Until this point, all the results presented in this section focus on initial conditions including only $n=2$ mode perturbation. Numerical results suggest that both the qualitative and some quantitative features discussed so far in this section also apply to dynamics with different symmetries. Here I further extend my analysis to a breakup dynamics with a $3$-fold symmetry. I impose the initial perturbation with one single $n=3$ vibrational mode. The same dynamics as in the $n=2$ mode perturbation is observed. For a range of initial conditions, the interface develops sharp tips followed by water fingers. In this case, the evolution of the tip curvature is qualitatively the same as that in Fig. \ref{fig:kmin_kmax} (b) (black open circles). The extreme tip curvature $k_{min}$ diverges as the initial condition varies. Here I fix the initial amplitude $A_3=0.01$ and vary the phase $\Omega_3$. The divergence of $k_{min}$ is fitted into a power function $(\Omega_3-\Omega_3^*)^\alpha$. The same value of $\alpha=-2.5$ obtained in the previous case with one single $n=2$ mode perturbation is used here to determine the threshold value $\Omega_3^*=0.27\pi$. This threshold value is then used to plot Fig. \ref{fig:n3} (a). It shows that the divergence of $k_{min}$ in the case of $n=3$ mode perturbation agrees with the exponent $\alpha=-2.5$ found earlier. However, starting with initial phase at the threshold value, as expected, the interface evolves into a coalescence (inset of Fig. \ref{fig:n3} (a)) before the tip curvature diverges and again this curvature singularity cannot be realized. In addition, for a phase space trajectory close to the curvature singularity, as shown in Fig. \ref{fig:n3} (b), the tip curvature and velocity deviate from the scalings dominated by the average collapse as the singularity is approached. Their divergences close to the formation of $k_{min}$ as functions of $(t_c-t)$ are described by the same exponents found earlier, i.e., $-0.8$ and $-0.4$ respectively. When the initial condition includes two modes with co-prime mode numbers (such as $n=2$ and $3$), the outcomes are more complicated and will be studied in future work.

\section{\label{section:discussion}Discussion}

\subsection{\label{subsection:physicaleffects}Other physical effects on curvature singularity}

As shown in previous section, both the curvature and velocity diverge as the curvature singularity is approached. The high velocity and high curvature suggest that there is a competition between the effect of inertia (high velocities) and other physical effects such as surface tension and viscosity that are ignored in my current model but may become important when the characteristic length scale of the problem is small (high curvatures). In addition, when the characteristic velocity increases to a value comparable to the sound speed in air, the effect of the compressibility of air also needs to be taken into account.
 
The significance of surface tension, viscosity and air compressibility are quantified by three dimensionless numbers as mentioned earlier: Weber number $We$, Reynolds number $Re$, and Mach number $Ma$ respectively. The Weber number $We$ and Reynolds number $Re$ describes the relative effects of inertia with respect to  surface tension and viscosity respectively, while the Mach number $Ma$ measures the importance of air compressibility. Their definitions are given by equation (\ref{eqn:re_we_ma}).These three effects will be important when the corresponding numbers $We$, $Re$ and $Ma$ reach unity.

In this section, I will first discuss the three effects on the formation of the curvature singularity by providing a scaling argument. Then I will show that if the curvature singularity were to form in experiments, it will be regularized by air viscosity and compressibility. 

First I discuss how the three effects affect the formation of the curvature singularity. Those three dimensionless numbers $We$, $Re$ and $Ma$ scale with the tip curvature $k$ and velocity ${\bf u}$ differently as $We\sim |{\bf u}|^2/|k|$, $Re\sim |{\bf u}|/|k|$, and $Ma\sim |{\bf u}|$. Fig. \ref{fig:k_v} shows that as the tip evolves towards the extreme curvature $k_{min}$, both the tip curvature $k$ and the speed of the tip point $|{\bf u}|$ diverge. Initially, they scale as $\bar{R}^{-1}$ when the perturbation amplitude is small compared with the average size of the cross-section. In this case, the Weber number $We$ and Mach number $Ma$ increase while the Reynolds number $Re$ remains approximately constant. When the tip curvature approaches $k_{min}$, the divergence of $k$ can be described by a power function of $(t_c-t)$ with an exponent approximately $-0.8$. Using the same threshold value $t_c$, the divergence of $|{\bf u}|$ is approximated as a power function of $(t_c-t)$ with an exponent around $-0.4$. Thus one should expect that as the curvature singularity is approached, the Weber number $We$ will stay approximately constant, the Reynolds number $Re$ will approach zero ($Re\sim (t_c-t)^{0.4}$), and the Mach number $Ma$ will diverge ($Ma\sim (t_c-t)^{-0.4}$). As a result, if the Weber number is relatively large initially, the effect of surface tension will remain marginal, while the air viscosity and compressibility will both be significant as the tip curvature diverges. 

If the curvature singularity were to form in experiments, since both $Re^{-1}$ and $Ma$ diverge with the same exponent, which one matters first depends on their pre-factors in the power-law scalings. Using the data attained from the simulation for initial condition $A_2=0.01$ and $\Omega_2=0.699\pi$ (Fig. \ref{fig:k_v} (a) (b), corresponding to the maximum $k_{min}$ for $A_2=0.01$), I obtain that
\begin{eqnarray}
|k|&=&0.07\times(t_c-t)^{-0.8}\label{eqn:kscaling}\\
|{\bf u}|&=&2.4\times(t_c-t)^{-0.4}
\label{eqn:uscaling}
\end{eqnarray}
(Here I focus on the scalings as the curvature singularity is approached. Specifically, in this case here ($A_2=0.01$, $\Omega_2=0.699\pi$), to obtain the pre-factors, I use data that correspond to $|k|>1000$ (to exclude the initial divergence dominated by average collapse) but exclude the last 5 points at the left end of each curve in Fig. \ref{fig:k_v} (b) where the curvature divergence slows down towards the emergence of $K_{min}$.)

For underwater air bubble breakup experiments \cite{keim06, keim11}, using scales estimated in Section \ref{section:problem}, I get
\begin{eqnarray}
We&=&70\\
Re^{-1}&=&0.0035\times(t_c-t)^{-0.4}\\
Ma&=&0.0035\times(t_c-t)^{-0.4}
\end{eqnarray}

Thus, with similar pre-factors in $Re^{-1}$ and $Ma$, both air viscosity and compressibility will be important in regularizing the curvature singularity in experiments. The crossover length scale of the average radius of the cross-section for which both $Re$ and $Ma$ hit unity is around $0.6\mu m$. For a different initial condition, those pre-factors are still within the same order of magnitude (although in some cases they may be different by up to around a factor of $10$), in which case one still expect both air viscosity and compressibility to be significant if the singularity were to form in bubble breakup experiments. 

However, it is noted that based on an argument that a smaller perturbation amplitude leads to a smaller length scale (a smaller average neck radius which is comparable to the perturbation amplitude) when the dynamics deviates from the average collapse and enters the region where $|k|$ and $|{\bf u}|$ are described by the above scalings (\ref{eqn:kscaling}) and (\ref{eqn:uscaling}), one may expect that the effect due to the compressibility of the gas flow dominates viscous effect in the limit that the perturbation amplitude goes to zero. Limited by the dynamic range of initial perturbation amplitudes, the data I have here cannot conclusively address such a possibility but it is worth pursuing in future investigations.

For experiments studying the collapse of a non-axisymmetric, impact-created air cavity in water \cite{enriquez10, enriquez11, enriquez11_arxiv}, the initial length scale ($\sim cm$) is larger compared with bubble breakup experiments while the initial velocity scale there is similar ($\sim m/s$). Hence for cavity collapse experiments, if the curvature singularity were to form, air compressibility will be the major mechanism to regularize the curvature singularity. The crossover length scale of the average radius of the cross-section for $Ma$ to hit unity is around $50\mu m$.

Another thing to show briefly here is that the effect of air compressibility may come into play even before the nonlinearity becomes strong if the perturbation amplitude is small. The dynamics is expected to be nonlinear when the average radius $\bar{R}$ is comparable to the initial perturbation amplitude $A_2$. At that moment, the non-dimensionalized velocity scale $|{\bf u}|$ is approximately the order of $1/\bar{R}\approx1/A_2$. Thus the Mach number is approximated as $Ma\approx u_0|{\bf u}|/C_g$. $u_0=0.5m/s$ is the initial velocity scale estimated from bubble breakup experiments (Section \ref{section:problem}) and $C_g=340m/s$ is the speed of sound in air. Thus $Ma\approx 1/(680\times A_2)$  when the dynamics is expected to be strongly nonlinear. According to the estimate here, the Mach number will reach unity just due to the initial growth dominated by the average collapse ($|{\bf u}|\sim \bar{R}^{-1}$), before $\bar{R}$ reaches $A_2$ if $A_2<1/680\approx 0.0015$. For initial perturbation amplitude smaller than this value, the air compressibility will be important even before the dynamics becomes strongly nonlinear. In bubble breakup experiments, this non-dimensionalized number $A_2=0.0015$ corresponds to a dimensional perturbation amplitude around $0.4\mu m$ based on the estimate in Section \ref{section:problem}.

\subsection{\label{subsection:experiments}Connection with experiments}
Here I discuss some possible connections with experiments. Dynamics closer to the curvature singularity can be obtained by imposing an initial condition with a smaller perturbation amplitude. However, starting with a smaller perturbation amplitude will make experimental observations of the sharp tips difficult because the average size of the bubble neck cross-section at the time the sharp tips form is comparable to the perturbation amplitude. Starting directly with a cross-section closer to a shape with sharp tips may provide a way to circumvent this difficulty. This can be achieved by either starting with a largely distorted cross-section shape, or using the same perturbation amplitude but a higher vibrational mode $n$ for the initial single mode perturbation. Experiments using initial perturbations with higher mode number $n$ (with $n$ up to 20) have been done by Enriquez \emph{et al.} \cite{enriquez10, enriquez11, enriquez11_arxiv}. Their experiments focus on the collapse of a non-axisymmetric cavity created by the impact of a disk. During the collapse, cross-section shapes with sharp tips are observed in these experiments for some initial conditions \cite{enriquez10, enriquez11_arxiv} but the dynamics of those sharp tips hasn't been measured quantitatively.

\section{Conclusion}
I have investigated numerically the asymmetric bubble breakup focusing on the case when the nonlinear interaction is strong. Boundary integral simulation results showed that starting with one single vibrational mode perturbation, for a continuous range of initial conditions, the dynamics organizes itself into a near singular state. I showed that previously found coalescence modes of breakups \cite{turitsyn09} are interspersed with cusp-like modes of breakups in which the air-water interface develops sharp tips that are often followed by re-entrant water fingers. The formation of the sharp tips corresponds to a phase space evolution controlled by a saddle-node. Namely, the air-water interface first evolves towards a shape with sharp tips whose radii of curvature are much smaller than the average neck radius, and then evolves away from it. Along a continuous curve of threshold values of initial conditions, the interface appears to evolve into a finite-time curvature singularity by developing sharp tips with infinite curvatures. However, starting with initial conditions on that curve, the interface actually evolves into a coalescence and the curvature singularity is pre-empted by coalescence. Numerical results further suggest that the curvature singularity can only be realized with vanishingly small perturbation amplitude. In this case, the formation of the sharp tips can be interpreted as a weakly first-order
transition which becomes second-order, corresponding to the formation of a finite-time curvature singularity, in the limit that the initial perturbation amplitude approaches zero. For a phase space trajectory close to the curvature singularity, as the singularity is approached, the tip curvature $k$ diverges approximately as $|k|\sim (t_c-t)^{-0.8}$ and the tip speed $|{\bf u}|$ diverges approximately as $|{\bf u}|\sim (t_c-t)^{-0.4}$. According to the scalings, both the viscous drag and the compressibility of air will be significant if such a curvature singularity were to form in experiments.  

\begin{acknowledgments}
I thank Konstantin Turitsyn and Wendy Zhang for their insight and support. I also thank Justin Burton, Daniel Herbst, William Irvine, Nathan Keim, Sidney Nagel, Samuel Oberdick, Robert Rosner, and Laura Schmidt for many stimulating discussions. This work was supported by NSF CBET-0967282 (PI Wendy W. Zhang).
\end{acknowledgments}

\appendix*

\section{Numerical implementation with boundary integral method}

Here I introduce briefly the simulation method I used for completeness. In the numerical scheme, the air-water interface $S$, starting as a Jordan curve (simple and closed), is discretized into $N$ boundary elements $E_i$ ($i = 1, ..., N$) separated by $N$ nodal points $\bf{x_i}$ ($i=1, ... , N$). Each boundary element is then represented by a line segment, a straight line connecting two adjacent nodal points (except in generating initial data, where cubic splines are used). The normal velocity $u_{\perp, i}^E$ and potential $\phi^E_i$ along the $i$th element $E_i$ are constant and the potential $\phi^E_i$ takes the arithmetic average of the potentials at the two end points $\bf{x_i}$ and $\bf{x_{i+1}}$ of the element $E_i$. With this discretization, equation (\ref{integraleqn}) can be schematically rewritten as:
\begin{equation}\label{integraleqndisc}
\frac{1}{2}{\phi^E}=-SLu^E_{\perp} + DL\phi^E
\end{equation}
where $SL$ and $DL$ are $N$ by $N$ matrices whose elements are represented by integrals along boundary elements and thus only depend on the interface shape, the Green's function and its normal derivative. $u^E_{\perp}$ is an $N$ by 1 vector with each element representing the normal velocity on corresponding boundary element $E_i$.

After the initial data are specified by the expansions in ${\cal R}$ and ${\cal V}$ (equation (\ref{eqn:Rexpansion}) and (\ref{eqn:Vexpansion})), the interface shape $S$ and the velocity along it are computed from equation (\ref{RVdefine}). Then the velocity potential $\phi$ can be solved numerically according to equation (\ref{integraleqndisc}). In this step, the elements in $SL$ and $DL$ are calculated using cubic splines (based on arc-length along the interface) between nodal points, and the corresponding integrals are performed using Gaussian quadratures over segments connecting adjacent splines' midpoints. 

Once the initial data, position $\bf{x_i}$ and potential $\phi_i$ at each nodal point, are obtained, the main simulation procedure below is similar to the one described by Pozrikidis \cite{pozrikidis_book}. The air pressure $p_{air}$ is adjusted at each time step to ensure the prescribed areal flux. The main simulation procedure is described briefly as follows:
\begin{enumerate}
\item{For each instance in time, the position and potential at the middle point of each line element $E_i$ is calculated by taking the arithmetic average of the data from the element's two end points. }
\item{Each element in the matrices $SL$ and $DL$ in equation (\ref{integraleqndisc}) is calculated using 20th order Gaussian quadrature. After that, equation (\ref{integraleqndisc}) is solved for the normal velocity $u^E_{\perp}$ at the middle point of every line element. Note here that, the air pressure term $p_{air}(t)$ in equation (\ref{bc:dynamic}) will shift the velocity potential along the interface by a constant which is only a function of time, and thus change the normal velocity solved from equation (\ref{integraleqndisc}). In this step, taking into account the effect of $p_{air}$, a constant value $\Delta\phi(t)$ is added to the potential $\phi^E$  and the value of $\Delta\phi(t)$ is determined by making the implosion areal flux $2\pi Q=\oint_Su_\perp ds$ the same as its prescribed value. Here $\oint_S$ means integral along the air-water interface.}
\item{At each nodal point ${\bf x_i}$, the normal velocity is calculated by taking the average of the normal velocities at the middle points of ${\bf x_i}$'s adjacent line elements $E_{i-1}$ and $E_i$, weighted by the inverses of the lengths of the two line elements. The tangential velocity, surface normal and tangential directions at the $i$th nodal point $\bf{x_i}$ are calculated by taking a finite difference using data from 3 points ($\bf{x_{i-1}}$, $\bf{x_i}$ and $\bf{x_{i+1}}$).}
\item{Then the potential $\phi_i$ is updated using the stress balance condition (\ref{bc:dynamic}), and the surface shape ${\bf x_i}$ is updated using the kinematic boundary condition (\ref{bc:kinematic}). The potential $\phi_i$ is updated using the stress balance condition (\ref{bc:dynamic}). The value of $p_{air}$ is determined later as in step $2$ to be consistent with a prescribed areal flux $2\pi Q$. }
\end{enumerate}
The above steps are repeated until one of the stop criteria mentioned in Section \ref{section:method} is met.

\nocite{*}
%


\end{document}